\documentclass[twocolumn,superscriptaddress,nofootinbib]{revtex4-1}

\usepackage[utf8]{inputenc} 
\usepackage[T1]{fontenc}    
\usepackage{hyperref}       
\usepackage{url}            
\usepackage{booktabs}       
\usepackage{amsfonts}       
\usepackage{nicefrac}       
\usepackage{microtype}      
\usepackage{lipsum}
\usepackage{amsmath}
\usepackage{graphicx}
\usepackage{xcolor}         
\definecolor{midgreen}{rgb}{0.52, 0.73, 0.4}

\usepackage{float}
\DeclareMathOperator{\sech}{sech}
\newcommand{\Lagr}{\mathcal{L}}

\begin{document}

\title{Hot and dense quark-gluon plasma thermodynamics from holographic black holes}

\author{Joaquin Grefa}
\affiliation{Physics Department, University of Houston, Houston TX 77204, USA}
\author{Jorge Noronha} 
\affiliation{Illinois Center for Advanced Studies of the Universe, Department of Physics, University of Illinois at Urbana-Champaign, Urbana, IL 61801, USA}

\author{Jacquelyn Noronha-Hostler}
\affiliation{Illinois Center for Advanced Studies of the Universe, Department of Physics, University of Illinois at Urbana-Champaign, Urbana, IL 61801, USA}

\author{Israel Portillo}
\affiliation{Physics Department, University of Houston, Houston TX 77204, USA}

\author{Claudia Ratti}
\affiliation{Physics Department, University of Houston, Houston TX 77204, USA}

\author{Romulo Rougemont}
\affiliation{Departamento de F\'{i}sica Te\'{o}rica, Universidade do Estado do Rio de Janeiro,
Rua S\~{a}o Francisco Xavier 524, 20550-013, Maracan\~{a}, Rio de Janeiro, Rio de Janeiro, Brazil}

\date{\today}

\begin{abstract}
We present new results on the equation of state and transition line of hot and dense strongly interacting QCD matter, obtained from a bottom-up Einstein-Maxwell-Dilaton holographic model. We considerably expand the previous coverage in baryon densities in this model by implementing new numerical methods to map the holographic black hole solutions onto the QCD phase diagram. We are also able to obtain, for the first time, the first-order phase transition line in a wide region of the phase diagram. Comparisons with the most recent lattice results for the QCD thermodynamics are also presented.
\end{abstract}

\maketitle 

\section{Introduction}

Significant efforts are underway to search for the quantum chromodynamics (QCD) critical point and subsequent first-order phase transition line at medium- to low-beam energies \cite{Bzdak:2019pkr}. Ongoing experiments such as the phase II of the beam energy scan at the Relativistic Heavy Ion Collider (RHIC), including a fixed-target program running at $\sqrt{s_{\textrm{NN}}}=3-7.7$ GeV \cite{STARnote,Cebra:2014sxa} and HADES at the GSI, with $\sqrt{s_{\textrm{NN}}}=1-3$ GeV \cite{Galatyuk:2014vha}, are currently looking for the QCD critical point. Additionally, next generation experiments such as FAIR at the GSI ($\sqrt{s_{\textrm{NN}}}=2.9-4.9$ GeV)
\cite{Friese:2006dj,Tahir:2005zz,Lutz:2009ff,Durante:2019hzd} and NICA in Dubna ($\sqrt{s_{\textrm{NN}}}=3-5$ GeV) \cite{Kekelidze:2017tgp,Kekelidze:2016wkp} are being built to precisely determine the QCD equation of state (EOS) and the properties of the strongly interacting quark-gluon plasma (QGP) at large baryon densities.
Relevant observables in this quest include fluctuations of conserved charges \cite{Stephanov:2008qz,Stephanov:2011pb,Bellwied:2018tkc,Adamczewski-Musch:2020slf,Adam:2020unf,Bellwied:2019pxh,Alba:2020jir,Mroczek:2020rpm}, flow \cite{Kardan:2018hna}, and particle yields \cite{Adamczewski-Musch:2019byl}. For recent reviews see Refs. \cite{Ratti:2018ksb,Bzdak:2019pkr}.
 
In order to simulate the evolution of heavy-ion collisions at low collision energies, the EOS is needed at large baryon chemical potential $\mu_B$. First principle lattice QCD calculations provide the EOS at  $\mu_B =0$ \cite{Borsanyi:2010cj,Borsanyi:2013bia,Bazavov:2014pvz}. However, due to the fermion sign problem \cite{Philipsen:2012nu}, it is not possible to directly calculate the EOS at finite densities. Nevertheless, one can reconstruct the EOS using susceptibilities calculated on the lattice through a Taylor series \cite{Allton:2002zi,Allton:2005gk,Borsanyi:2012cr,Bazavov:2017dus,DElia:2002tig,DElia:2016jqh,Bazavov:2017dus,Gunther:2016vcp,Borsanyi:2018grb,Bazavov:2020bjn,Parotto:2018pwx,Noronha-Hostler:2019ayj,Monnai:2019hkn,Everett:2020yty}, currently limited to $\mu_B/T\leq 2$ (where $T$ is the temperature). A new expansion has been proposed in Ref.\ \cite{Borsanyi:2021sxv}, which covers a much larger region of $\mu_B$ with high precision. Unfortunately, such an approach cannot cover the whole phase diagram, nor can it accurately capture critical behavior. Therefore, one must turn to alternative approaches to describe the matter created in low-energy collisions, and in the vicinity of a critical point. A promising effective theory should not only reproduce lattice QCD thermodynamics results where they are available, but also the QGP's nearly perfect fluid behavior \cite{Heinz:2013th} implied by current extractions of its transport properties from comparisons between model calculations and experimental data \cite{Bernhard:2016tnd,Bernhard:2019bmu}. To the best of our knowledge, the only effective model currently available in the literature that can simultaneously describe on a \textit{quantitative level} both equilibrium and near-equilibrium features of the strongly coupled QGP is the bottom-up non-conformal Einstein-Maxwell-Dilaton (EMD) holographic model proposed by some of us in Ref.\ \cite{Critelli:2017oub}. This model, which is able to quantitatively describe the high-order baryon susceptibilities obtained on the lattice and the nearly perfect fluid behavior of the QGP, is built up on the general reasoning originally laid down in the seminal works of \cite{Gubser:2008yx,DeWolfe:2010he,DeWolfe:2011ts}, based on a phenomenological approach of the well-known gauge/gravity duality \cite{Maldacena:1997re,Gubser:1998bc,Witten:1998qj,Witten:1998zw}. Some previous holographic approaches focusing on qualitative aspects of the strongly coupled QGP can be seen e.g. in Refs.\ \cite{Kovtun:2004de,CasalderreySolana:2011us,Ficnar:2010rn,Ficnar:2011yj,Ficnar:2012yu,Finazzo:2013efa,Finazzo:2014cna,Rougemont:2015ona,Rougemont:2015wca,Finazzo:2015xwa,Rougemont:2017tlu,Rougemont:2018ivt,Rougemont:2015oea,Finazzo:2016mhm,Critelli:2016cvq,Rougemont:2020had,Knaute:2017opk,Li:2017ple}.

The construction of the EMD model of Ref.\ \cite{Critelli:2017oub} mainly differs from the earlier developments of \cite{Gubser:2008yx,DeWolfe:2010he,DeWolfe:2011ts} by the fact that the old lattice data used in those previous holographic works to fix the free parameters of the model are, nowadays, known not to be quantitatively accurate. On the other hand, Ref.\ \cite{Critelli:2017oub} makes use of state-of-the-art lattice QCD results at $\mu_B=0$ as first principles inputs from QCD to fix the free parameters of the EMD model, as we are going to review in section \ref{sec:fix}. Moreover, as discussed in Appendix A of Ref. \cite{Critelli:2017oub}, in the EMD model constructed in Refs. \cite{DeWolfe:2010he,DeWolfe:2011ts} four different dimensionful scales were introduced to express the temperature, baryon chemical potential, entropy density and baryon charge density in physical units, while in QCD there is just one dimensionful scale, $\Lambda_{\textrm{QCD}}$. Therefore, also in the EMD model constructed in Ref. \cite{Critelli:2017oub} there is a single dimensionful scale, $\Lambda$, which is used to express any physical observable in physical units (see section \ref{sec:fix}).

The EMD model of Ref.\ \cite{Critelli:2017oub} predicted a critical point in the QCD phase diagram at $T\sim 89$ MeV and $\mu_B\sim 724$ MeV. However, even though holographic calculations at finite chemical potentials are not affected by the fermion sign problem, numerical calculations at very large $\mu_B$ are still quite challenging in this approach, which was the reason why in Ref.\ \cite{Critelli:2017oub} some of us were still unable to locate the line of first-order phase transition in the region beyond the critical point, as we will discuss in detail in the present work. Here we considerably expand our previous results \cite{Critelli:2017oub} by overcoming most numerical difficulties and providing our equation of state over a broad range in temperature (2 MeV $\leq T\leq$ 550 MeV) and baryon chemical potential (0 $\leq\mu_B\leq$ 1100 MeV). By mapping out the phase diagram of our model within this unprecedentedly large region in the $(T,\mu_B)$ plane, we finally locate the first-order phase transition line beyond the critical point of our model originally calculated in Ref.\ \cite{Critelli:2017oub}. Moreover, with the filtering scheme developed in the present work to smooth out numerical noise, we were also able to calculate the physical observables on top of the phase transition regions, which was something we were unable to do at the time of publication of Ref.\ \cite{Critelli:2017oub}. Furthermore, we also present in this work comparisons between our results and the latest lattice data from Ref.\ \cite{Borsanyi:2021sxv}.

The paper is organized as follows. In Sections \ref{sec:EMD} and \ref{sec:EoM} we review some of the main aspects of the bottom-up EMD model proposed in Ref.\ \cite{Critelli:2017oub}, which are necessary in the implementation of our new numerical developments presented in detail in Section \ref{sec:results}. Also, in Section \ref{sec:results} we present our results for the thermodynamic quantities of the strongly coupled QGP, largely extending the range of values of $\mu_B$ covered in the phase diagram of the EMD model, which allows us to locate the first-order phase transition past  the critical point originally obtained in Ref.\ \cite{Critelli:2017oub}. In Section \ref{sec:conclusion} we present our conclusions and future perspectives in face of the results discussed here. In the present work we employ natural units $\hbar=c=k_B=1$ and a mostly plus metric signature.

\section{The Holographic EMD Model}
\label{sec:EMD}

Through the holographic gauge/gravity correspondence developed in string theory, calculations of physical observables in a strongly coupled quantum non-Abelian gauge theory in (flat) four dimensions can be performed by solving the classical equations of motion of a higher dimensional theory of gravity in asymptotically Anti-de Sitter (AdS) spacetimes. In the present work we employ a five-dimensional bottom-up EMD model defined by the following action \cite{DeWolfe:2010he,Critelli:2017oub}
\begin{eqnarray} \label{eq:action}
    S&=&\int_{\mathcal{M}_5} d^5 x \Lagr = \frac{1}{2\kappa_{5}^{2}}\int_{\mathcal{M}_5} d^{5}x\sqrt{-g}\times
    \nonumber\\
    &\times&\left[R-\frac{(\partial_\mu \phi)^2}{2}-V(\phi)-\frac{f(\phi)F_{\mu\nu}^{2}}{4}\right],
\end{eqnarray}
where $\kappa_5^2\equiv 8\pi G_5$ and $G_5$ is the five-dimensional Newton's constant. The EMD action (\ref{eq:action}) comprises three bulk fields in five dimensions: the metric $g_{\mu\nu}$, a real scalar called the dilaton field $\phi$, and a Maxwell field $A_\mu$. Additionally, $R$ is the Ricci scalar and $F_{\mu\nu} = \nabla_\mu A_\nu - \nabla_\nu A_\mu$. We set the asymptotic AdS$_5$ radius $L$ to unity and introduce as a free parameter in its place an energy scale $\Lambda$, which is going to be fixed together with $\kappa_5^2$ in Section \ref{sec:fix}. The single energy scale $\Lambda$ expressed in MeV will be used to write in physical units the gauge theory observables originally calculated in terms of inverse powers of $L$ on the gravity side of the holographic gauge/gravity correspondence.

We note that Eq.\ (\ref{eq:action}) is the simplest five-dimensional action that can holographically produce a phenomenologically realistic QCD-like effective theory in four dimensions at finite temperature and chemical potential. In what follows, we review some of the main aspects of the EMD model already presented in detail in Ref.\ \cite{Critelli:2017oub}, since they are important for the new numerical procedure we develop in the present work, which shall be discussed in Sections \ref{sec:num} and \ref{sec:EOS}.

The five-dimensional metric $g_{\mu\nu}$ is dual to the stress-energy tensor of the four dimensional quantum gauge theory and the extra holographic direction may be interpreted as a geometrization of the energy scale of the renormalization group flow of the gauge theory \cite{deBoer:1999tgo}. The dilaton field is used in the present setup to break the conformal invariance of the theory, with its potential $V(\phi)$ (and also the free parameters $\kappa_5^2$ and $\Lambda$) being engineered in a very specific way such as to emulate the behavior of the QGP in equilibrium, as inferred from lattice QCD calculations at $\mu_B=0$. The Maxwell field is employed here to introduce the effects associated with a finite baryon chemical potential, which is done by tuning the coupling function $f(\phi)$ in order to have the holographic baryon susceptibility matching the corresponding lattice QCD result also at $\mu_B=0$. Therefore, as we are going to review in Section \ref{sec:fix}, all the free parameters of our EMD model are fixed by lattice QCD inputs at zero net baryon density. Consequently, all the observables calculated at nonzero $\mu_B$, besides all of those computed at $\mu_B=0$ which were not used to fix the free parameters of the EMD action, follow as \emph{bona fide} predictions of our holographic model. 

We are interested here in five-dimensional, non-rotating, translationally invariant, spatially isotropic, and charged black hole backgrounds in thermodynamic equilibrium. In this case, the EMD fields are described by the following general Ansatz \cite{DeWolfe:2010he}
\begin{equation}
\begin{array}{rcl} \label{eq:ansatz}
     ds^2 &=& e^{2A(r)}[-h(r)dt^2+d\vec{x}^2]+\frac{e^{2B(r)}dr^{2}}{h(r)}, \\
     \phi &=& \phi(r), \\
     A &=& A_{\mu}dx^{\mu}=\Phi(r)dt,
\end{array}
\end{equation}
where $r$ is the holographic coordinate. The radial position of the black hole event horizon is given by the largest root of $h(r_{H})$=0 and the boundary of the asymptotically AdS$_5$ geometry lies at $r\rightarrow \infty$.  
The equations of motion (EoM) can be readily obtained
\begin{eqnarray}\label{eq:EoM1}
&&\phi''(r)+\left[\frac{h'(r)}{h(r)}+4A'(r)-B'(r)\right]\phi'(r)+
\\
&-&\frac{e^{2B(r)}}{h(r)}\left[\frac{\partial V(\phi)}{\partial\phi}-\frac{e^{-2[A(r)+B(r)]}\Phi'(r)^{2}}{2}\frac{\partial f(\phi)}{\partial\phi}\right]=0,
\nonumber
\end{eqnarray}
\begin{equation}\label{eq:EoM2}
\Phi''(r)+\left[2A'(r)-B'(r)+\frac{d[\ln{f(\phi)}]}{d\phi}\phi'(r)\right]\Phi'(r)=0,
\end{equation}
\begin{equation}\label{eq:EoM3}
A''(r)-A'(r)B'(r)+\frac{\phi'(r)^{2}}{6}=0,
\end{equation}
\begin{equation}\label{eq:EoM4}
h''(r)+[4A'(r)-B'(r)]h'(r)-e^{-2A(r)}f(\phi)\Phi'(r)^{2}=0,
\end{equation}
\begin{eqnarray}
\label{eq:EoM5}
&&h(r)[24A'(r)^{2}-\phi'(r)^{2}]+6A'(r)h'(r)+
\nonumber\\
&&+2e^{2B(r)}V(\phi)+e^{-2A(r)}f(\phi)\Phi'(r)^{2}=0,
\end{eqnarray}
with Eq.\ \eqref{eq:EoM5} being a constraint. Since the background function $B(r)$ has no dynamics, one may employ a gauge choice where $B(r)=0$ in order to simplify the numerical calculations, as we are going to do in a moment.

The equation of motion for $\Phi(r)$ can be integrated to obtain the conserved Gauss charge $Q_{G}$ associated with the gauge field $A_{\mu}$,
\begin{equation}\label{eq:GaussCharge}
    Q_{G}(r)=f(\phi)e^{2A(r)-B(r)}\Phi'(r).
\end{equation}
From Eq.\ (\ref{eq:EoM4}) for the blackening function $h(r)$ another conserved charge is obtained: the Noether charge
\begin{equation}\label{eq:NoetherCharge}
    Q_{N}=e^{2A(r)-B(r)}\left[e^{2A(r)h'(r)-f(\phi)\Phi(r)\Phi'(r)}\right].
\end{equation}

\section{Numerical solutions to the EoM and Thermodynamic quantities}
\label{sec:EoM}

In order to solve the EoM numerically, we need to define a different set of coordinates which we call ``numerical coordinates'', in addition to the so-called ``standard coordinates'', which will be denoted with a tilde. Both sets of coordinates are defined in the gauge where $B(r)=0$. One may calculate the thermodynamic quantities such as entropy density and temperature from standard holographic formulas using the standard coordinates, in terms of which $\tilde{h}(\tilde{r}\to\infty)=1$, as usual. However, to numerically solve the EoM, it is necessary to rescale the standard coordinates to specify definite values for some of the Taylor coefficients in the near-horizon expansions of the EMD fields, as required in order to initialize the numerical integration of the differential equations \eqref{eq:EoM1} --- \eqref{eq:EoM4}. This rescaling is accomplished using the numerical coordinates, as we discuss next.

\subsection{Standard coordinates and thermodynamics}
The near-boundary, ultraviolet expansions of the EMD fields in the standard coordinates read \cite{DeWolfe:2010he,Critelli:2017oub}
\begin{equation}
    \begin{array}{rcl}
         \tilde{A}(\tilde{r})&=& \tilde{r}+O\left(e^{-2\nu\tilde{r}} \right),  \\
         \tilde{h}(\tilde{r})&=& 1+O\left(e^{-4\tilde{r}} \right),  \\
         \tilde{\phi}(\tilde{r})&=& e^{-\nu\tilde{r}}+O\left(e^{-2\nu\tilde{r}}\right),  \\
         \tilde{\Phi}(\tilde{r})&=& \tilde{\Phi}_{0}^{\textrm{far}}+ \tilde{\Phi}_{2}^{\textrm{far}}e^{-2\tilde{r}}+O\left(e^{-(2+\nu)\tilde{r}}\right), 
    \end{array}
\end{equation}
where $\nu\equiv d-\Delta$, $d=4$ is the number of spacetime dimensions of the boundary gauge theory, $\Delta=(d+\sqrt{d^{2}+4m^{2}})/2$ is the scaling dimension of the gauge field theory operator dual to the dilaton $\phi(r)$ and $m$ is the mass of the dilaton field obtained from the dilaton potential (which will be specified in Section \ref{sec:fix}).

The temperature of the gauge theory fluid corresponds to the Hawking temperature of the black hole solution
\begin{equation}\label{eq:HawkingT}
T=\left.\frac{\sqrt{-g'_{\tilde{t}\tilde{t}}g^{\tilde{r}\tilde{r}'}}}{4\pi}\right|_{\tilde{r}=\tilde{r}_{H}} \!\!\!\!\!\!\!\!\!\!\!\!\Lambda=\frac{e^{\tilde{A}(\tilde{r}_{H})}}{4\pi}|\tilde{h}'(\tilde{r}_{H})|\Lambda,
\end{equation}
where we already introduced the energy scale $\Lambda$ so that Eq.\ \eqref{eq:HawkingT} gives the temperature of the QGP expressed in MeV. The entropy density of the boundary fluid is related to the area of the black hole event horizon, $A_H$, via the Bekenstein-Hawking formula \cite{Bekenstein:1973ur,Hawking:1974sw}
\begin{equation}\label{eq:HawkingS}
    s=\frac{S}{V}\Lambda^3=\frac{A_{H}}{4G_{5}V}\Lambda^3=\frac{2\pi}{\kappa_{5}^{2}}e^{3\tilde{A}(\tilde{r}_{H})}\Lambda^{3},
\end{equation}
where $V$ is the 3-dimensional spatial volume.
One can also obtain the baryon chemical potential of the system from the boundary value of the Maxwell field
\begin{equation}\label{eq:chemPot}
    \mu_{B}=\lim_{\tilde{r}\rightarrow\infty}\tilde{\Phi}(\tilde{r})\Lambda=\tilde{\Phi}_{0}^{\textrm{far}}\Lambda,
\end{equation}
whereas the baryon density is obtained from the boundary value of the radial momentum conjugate to the Maxwell field
\begin{equation}\label{eq:rhoB}
\rho_{B}=\lim_{\tilde{r}\rightarrow\infty}\frac{\partial\mathcal{L}}{\partial(\partial_{\tilde{r}}\tilde{\Phi})}\Lambda^{3}= \frac{Q_{G}(\tilde{r}\rightarrow\infty)}{2\kappa_{5}^{2}}\Lambda^{3}=-\frac{\tilde{\Phi}_{2}^{\textrm{far}}}{\kappa_{5}^{2}}\Lambda^3.
\end{equation}

\subsection{Thermodynamics in the numerical coordinates}
For numerically solving the EMD EoM, we consider Taylor expansions of the bulk fields near the black hole event horizon, $\sum_{n=0}^{\infty}X_{n}(r-r_{H})^n$, where $X={A,h,\phi,\Phi}$. We rescale the holographic coordinate $r$ so that $r_{H}=0$. The fact that the blackening function has a simple zero at the horizon leads to $h_{0}=0$. Also, $A_{0}=0$ can be fixed by rescaling the spacetime coordinates $(t,\vec{x})$ by a common factor, while $h_{1}=1$ can be arranged by rescaling only the time coordinate $t$. In addition, one must impose $\Phi_{0}=0$ for $\Phi dt$ to be well-defined, since $dt$ has infinite norm at the horizon. With the Taylor coefficients $h_{0}$, $h_{1}$, $A_{0}$, and $\Phi_{0}$ determined as aforementioned, the solutions to the set of equations \eqref{eq:EoM1} --- \eqref{eq:EoM5} via Taylor expansions can be parametrized by just two coefficients, namely the value of the dilaton field calculated at the horizon, $\phi_{0}$, and the derivative of the Maxwell field evaluated at the horizon, $\Phi_{1}$. Indeed, different choices for the pair of initial conditions $(\phi_0,\Phi_1)$ produce different black hole geometries, each of them corresponding to some definite thermal state of the gauge theory in equilibrium. The phase diagram of the model can be then populated in the $(T,\mu_B)$ plane by considering a large ensemble of different black hole solutions.

During the numerical integration of the equations of motion, we avoid the singularity at the horizon ($r_{H}=0$) by starting at a slightly shifted position, e.g $r_{\textrm{start}}\equiv 10^{-8}$. The boundary can be numerically parameterized by the value of the holographic coordinate $r$ at which the EMD fields have already reached their ultraviolet behavior corresponding to the AdS$_{5}$ geometry, which has a Ricci scalar of $R=-20$. For the vast majority of initial conditions considered in the present work, the corresponding black hole solutions satisfy this condition for $r\lesssim r_{\textrm{max}}=2$, which is then taken as the upper bound for the numerical integration of the EoM and can be used as a numerical parametrization of the boundary. However, for some initial conditions the dilaton only reaches the value of $10^{-5}$ for larger values of $r_{\textrm{max}}$ (such a small value of the dilaton is used as part of our algorithm to extract its leading ultraviolet coefficient close to the boundary, as discussed below); in such cases we simply set $r_{\textrm{max}}=10$.

The asymptotics of the EMD fields also imply the following bound for generating asymptotically AdS$_5$ solutions from the chosen values of the pair of initial conditions $(\phi_0,\Phi_1)$ \cite{DeWolfe:2010he,Critelli:2017oub}
\begin{equation}
    \Phi_{1}<\sqrt{-\frac{2V(\phi_{0})}{f(\phi_{0})}}\equiv \Phi_{1}^{\textrm{max}}(\phi_{0}).
\end{equation}

In the numerical coordinates, one can show that the ultraviolet behavior of the EMD fields is given according to \cite{DeWolfe:2010he,Critelli:2017oub}
\begin{equation}
\label{UVfields}
    \begin{array}{rcl}
         A(r)&=& \alpha(r)+O\left(e^{-2\nu\alpha(r)} \right),  \\
         h(r)&=& h_{0}^{\textrm{far}}+O\left(e^{-4\alpha(r)} \right),  \\
         \phi(r)&=& \phi_{A}e^{-\nu\alpha(r)}+O\left(e^{-2\nu\alpha(r)}\right),  \\
         \Phi(r)&=& \Phi_{0}^{\textrm{far}}+ \Phi_{2}^{\textrm{far}}e^{-2\alpha(r)}+O\left(e^{-(2+\nu)\alpha(r)}\right), 
    \end{array}
\end{equation}
where $\alpha(r)=A_{-1}^{\textrm{far}}r+A_{0}^{\textrm{far}}$. By calculating the constraint Eq.\ \eqref{eq:EoM5} at the boundary, one obtains $A_{-1}^{\textrm{far}}=1/\sqrt{h_{0}^{\textrm{far}}}$. Furthermore, by equating the conserved charge \eqref{eq:GaussCharge} evaluated at the boundary and at the horizon, one also finds that
\begin{equation}
\label{eq:Phi2}
\Phi_{2}^{\textrm{far}}=-\frac{\sqrt{h_0^{\textrm{far}}}}{2f(0)}f(\phi_{0})\Phi_{1}.
\end{equation}

For the kind of calculations we pursue here, we just need to obtain the behavior of a few ultraviolet expansion coefficients of the EMD fields near the boundary, namely $h_{0}^{\textrm{far}}$, $\Phi_{0}^{\textrm{far}}$, $\Phi_{2}^{\textrm{far}}$, and $\phi_A$. As discussed in Ref.\ \cite{Critelli:2017oub}, one may set $h_{0}^{\textrm{far}}=h(r_{\textrm{max}})$ and $\Phi_{0}^{\textrm{far}}=\Phi(r_{\textrm{max}})$, since the blackening function and the Maxwell field quickly reach their respective conformal values. From Eq.\ \eqref{eq:Phi2} one obtains $\Phi_{2}^{\textrm{far}}$, while $\phi_A$ can be reliably estimated by fitting the numerical solution for $\phi(r)$ using its ultraviolet asymptotics, $\phi_A e^{-\nu\alpha(r)}$, within the adaptive range $r\in[\phi^{-1}(10^{-3}),\phi^{-1}(10^{-5})]$. Notice that $\phi_A$ must be extracted from the comparison between the leading term in the analytic near-boundary expansion of the dilaton field and its full numerical result. Clearly the numerical solutions for the dilaton only converge to the corresponding ultraviolet asymptotics near the boundary, when the value of the dilaton approaches zero exponentially. The aforementioned adaptive region was defined after tests with several different initial conditions by considering the requirement that the numerical error defined as the difference between the numerical dilaton and its analytic leading order ultraviolet asymptotics is small when compared to the numerical value of the dilaton within the fitting region. When this requirement is satisfied, one can guarantee that $\phi_A$ is being reliably estimated. We have also considered different adaptive regions to extract the value of $\phi_A$, but always restricted to the requirement that this relative error must be small. The physical results remain unchanged as long as this requirement is met.

One can show that the thermodynamic variables \eqref{eq:HawkingT} --- \eqref{eq:rhoB} can be directly expressed in the numerical coordinates as follows \cite{Critelli:2017oub}
\begin{equation} \label{eq:T_bh}
    T=\frac{1}{4\pi\phi_{A}^{1/\nu}\sqrt{h_{0}^{\textrm{far}}}}\Lambda,
\end{equation}
\begin{equation} \label{eq:mu_bh}
    \mu_{B}=\frac{\Phi_{0}^{\textrm{far}}}{\phi_{A}^{1/\nu}\sqrt{h_{0}^{\textrm{far}}}}\Lambda,
\end{equation}
\begin{equation} \label{eq:s_bh}
    s=\frac{2\pi}{\kappa_{5}^{2}\phi_{A}^{3/\nu}}\Lambda^{3},
\end{equation}
\begin{equation} \label{eq:rho_bh}
    \rho_{B}=-\frac{\Phi_{2}^{\textrm{far}}}{\kappa_{5}^{2}\phi_{A}^{3/\nu}\sqrt{h_{0}^{\textrm{far}}}}\Lambda^{3}.
\end{equation}


\subsection{Fixing the free parameters of the EMD Model}
\label{sec:fix}

\begin{figure}[h]
    \centering
    \includegraphics[width=0.5\textwidth]{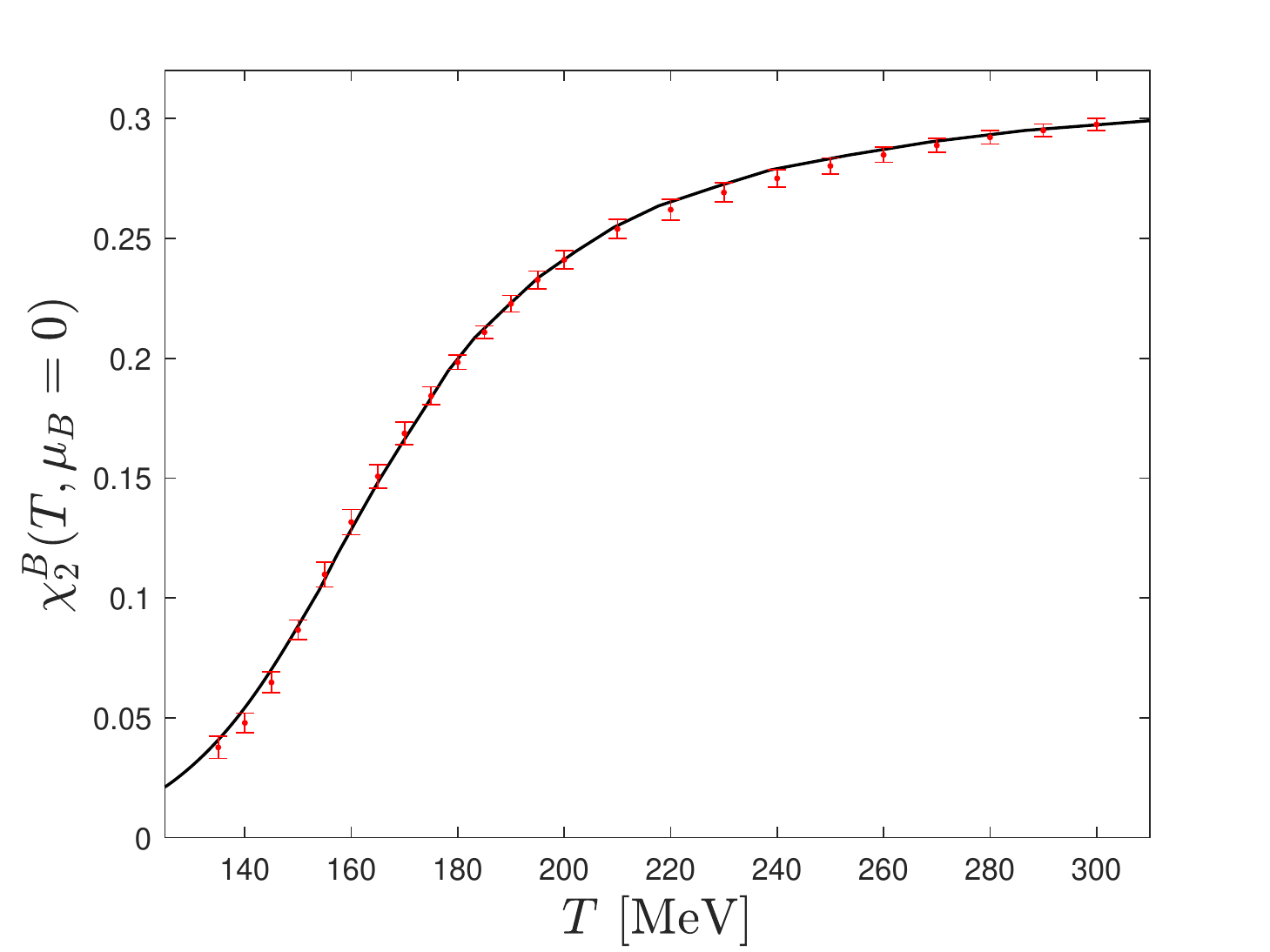}
    \caption{Results from the fitting of the holographic susceptibility (solid black curve) to the dimensionless second order baryon susceptibility $\chi_{2}^{B}(T,\mu_{B}=0)$ from lattice QCD \cite{Bellwied:2015lba}.}
    \label{fig:chi2_B0}
\end{figure}

\begin{figure}[h]
    \centering
    \includegraphics[width=0.5\textwidth]{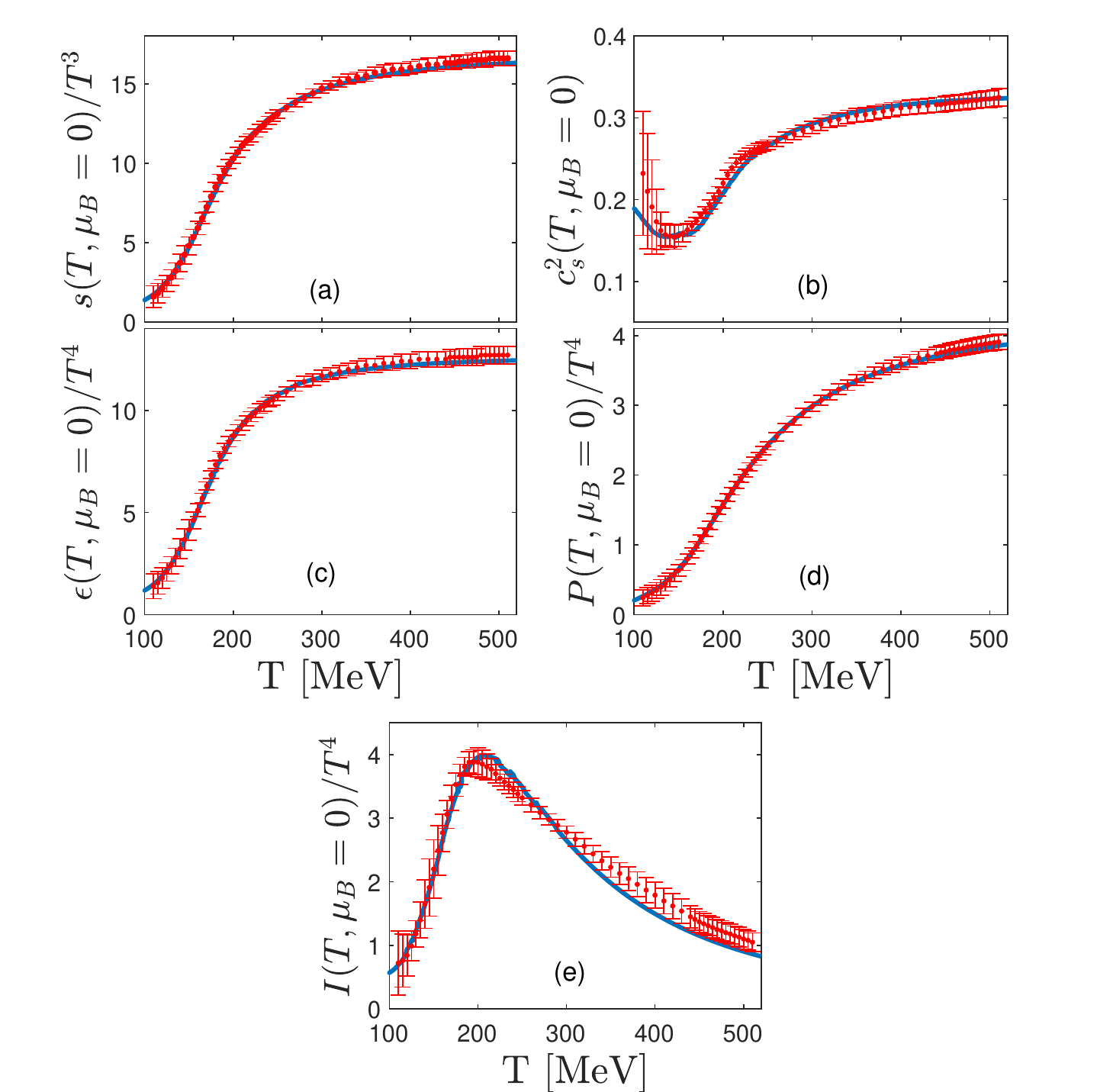}
    \caption{Thermodynamics at $\mu_{B}=0$. Lattice QCD results from Ref.\ \cite{Borsanyi:2013bia} (red points) are compared to the holographic model curves (blue lines): (a) entropy density, (b) speed of sound squared, (c) energy density $\epsilon$, (d) pressure $P$, and (e) trace anomaly $I=\epsilon-3P$.}
    \label{fig:matchingQCD}
\end{figure}

The free parameters of our EMD model, namely, $\kappa_5^2$, $\Lambda$, $V(\phi)$, and $f(\phi)$, are dynamically fixed by state-of-the-art lattice QCD inputs at $\mu_B=0$ with $2+1$ flavors and physical values of the quark masses. More specifically, $\kappa_5^2$, $\Lambda$, and $V(\phi)$ are fixed in order that the holographic equation of state at $\mu_B=0$ closely matches the corresponding lattice QCD results from Ref.\ \cite{Borsanyi:2013bia}, while $f(\phi)$ is fixed by requiring that the holographic second order baryon susceptibility, also calculated at $\mu_B=0$, closely matches the corresponding lattice result from Ref.\ \cite{Bellwied:2015lba}. In particular, at vanishing chemical potential it is possible to derive a holographic formula for the dimensionless second order baryon susceptibility, $\chi_{2}^B\equiv\partial^2(P/T^4)/\partial(\mu_B/T)^2$, which reads as follows \cite{DeWolfe:2010he,Rougemont:2015wca}
\begin{equation}
    \chi_{2}^B(\mu_{B}=0)=\frac{1}{16\pi^{2}}\frac{s}{T^{3}}\frac{1}{f(0)\int_{r_{H}}^{\infty}dr\ e^{-2A(r)}f(\phi(r))^{-1}},
\end{equation}
which is to be evaluated by setting the initial condition $\Phi_{1}$ to zero. In numerical calculations, we substitute $r_{H}\rightarrow r_{\textrm{start}}$ and $\infty\rightarrow r_{\textrm{max}}$.

In this way, the free parameters of our holographic EMD model are fixed as below,
\begin{equation}\label{eq:free_functions}
    \begin{array}{rcl}
         V(\phi)&=& -12\cosh(0.63\,\phi)+0.65\,\phi^{2}-0.05\,\phi^{4}+0.003\,\phi^{6},  \\
         \\
         \kappa_{5}^{2} &=& 8\pi G_{5}=8\pi(0.46), \qquad \Lambda=1058.83\, \textrm{MeV},  \\
         \\
         f(\phi) &=& \frac{\sech(c_{1}\phi+c_{2}\phi^{2})}{1+c_{3}}+\frac{c_{3}}{1+c_{3}}\sech(c_{4}\phi),
    \end{array}
\end{equation}
where $c_{1}=-0.27$, $c_{2}=0.4$, $c_{3}=1.7$, and $c_{4}=100$, with the corresponding fitting results displayed in Figs. \ref{fig:chi2_B0} and \ref{fig:matchingQCD}. As discussed in Ref.\ \cite{Critelli:2017oub}, the scaling dimension of the gauge theory operator dual to the dilaton field in our approach is $\Delta \approx 2.73$, which is a result \emph{implied} by dynamically matching, with our choice of the functional form of $V(\phi)$, the holographic equation of state to the corresponding state-of-the-art continuum extrapolated lattice QCD results evaluated at zero baryon density with 2+1 flavours and physical values of the quark masses. While one may follow \cite{Gubser:2008yx} and identify this scalar operator with $\mathrm{Tr}\,F^2$ in the gauge theory (which for $\Delta \approx 2.73$ would possess a large anomalous dimension), such formal identification is not rigorously needed to compute thermodynamic observables in a bottom-up approach.

We note that (an approximation for) the pressure can be easily calculated by integrating the entropy density with respect to the temperature,
\begin{equation}
    P(T, \mu_{B}=0)\approx \int_{T_{\textrm{low}}}^{T} dT \,s(T,\mu_{B}=0),
\end{equation}
where we take here $T_{\textrm{low}}=2$ MeV (this is the lowest value of temperature for the black hole solutions generated with the set of initial conditions considered in the present work, see Section \ref{sec:mapping}).

\section{Thermodynamics at finite chemical potential}
\label{sec:results}

With the results of Eqs. (\ref{eq:T_bh}) - (\ref{eq:rho_bh}), we can calculate many thermodynamic observables at finite temperature and baryon density. For instance, the internal and free energy densities at finite $\mu_{B}$ are, respectively,
\begin{eqnarray}
    \epsilon(s,\rho_{B})&=&Ts-P+\mu_{B}\rho_{B},
    \\
\label{eq:Free_energy}
    F(T,\mu_{B})&=&-P(T,\mu_{B})=\epsilon(s,\rho_{B})-Ts-\mu_{B}\rho_{B}
\end{eqnarray}
from which we can obtain the differential relations
\begin{eqnarray}
    d\epsilon(s,\rho_{B})&=&Tds+\mu_{B}d\rho_{B},
\\
    dF(T,\mu_{B})&=&-dP(T,\mu_{B})=-sdT-\rho_{B}d\mu_{B},
    \label{eq:dF}
\end{eqnarray}
so that at fixed $\mu_{B}$,
\begin{equation} \label{eq:diff_Pressure}
         dP(T,\mu_{B})|_{\mu_B}=sdT,
\end{equation}
and the square of the speed of sound at fixed $\mu_{B}$ reads,
\begin{equation} \label{eq:c2s0}
\!\tilde{c}_{s}^{2} = \left.\frac{dP}{d\epsilon}\right|_{\mu_{B}} \!\!\!\!\!\! 
         = \left(\frac{T}{s}\left.\frac{\partial s(T,\mu_{B})}{\partial T}\right|_{\mu_{B}}\!\!\!\!\!\!+\frac{\mu_{B}}{s}\left.\frac{\partial\rho_{B}(T,\mu_{B})}{\partial T}\right|_{\mu_{B}}\right)^{-1}\!\!\!\!\!\!. 
\end{equation}
\begin{figure*}[ht!]
    \centering
    \includegraphics[width=\textwidth]{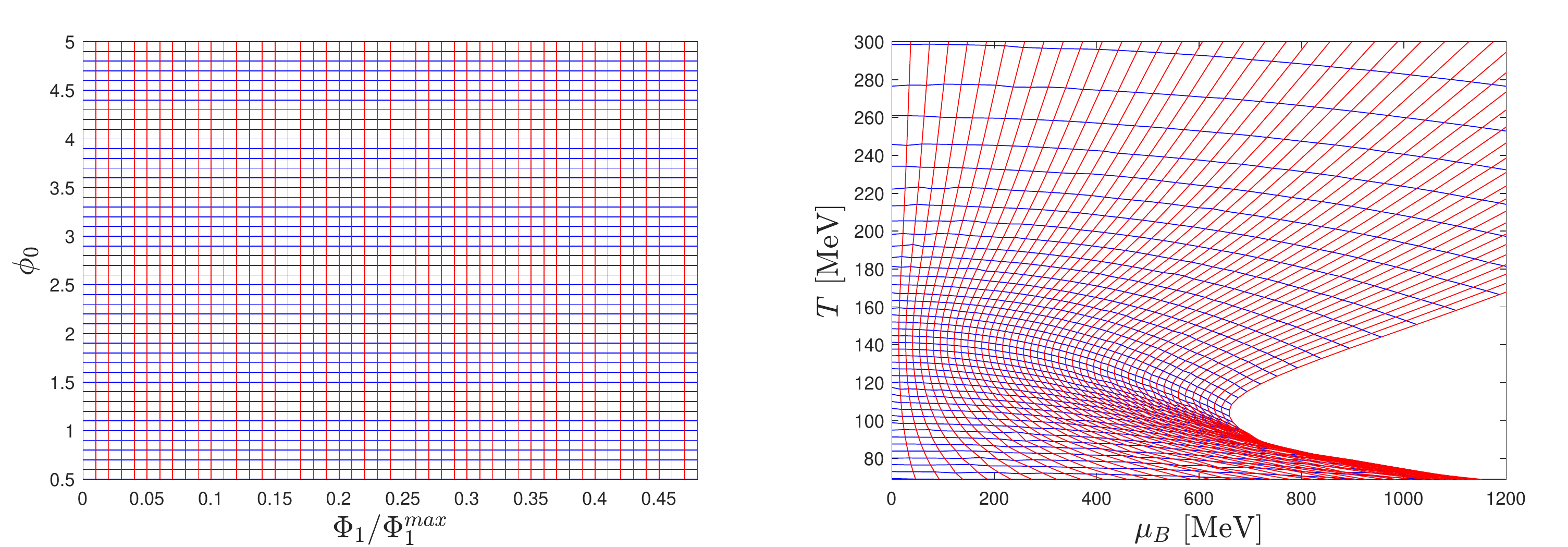}
    \caption{Mapping of an equally spaced, rectangular grid of initial conditions ($\phi_{0}$,$\Phi_{1}$) into an irregular grid of points in the ($T,\mu_{B}$) plane generated by the corresponding black hole solutions.}
    \label{fig:mapping}
\end{figure*}
Eq.\ \eqref{eq:c2s0} was used in Ref.\ \cite{Critelli:2017oub} to calculate the minimum of $\tilde{c}_{s}^{2}(T,\mu_B)$, which may be used as a ``transition line" characterizing the crossover region. However, although \eqref{eq:c2s0} is computationally simple to determine along trajectories at constant chemical potential, we note that a definition of the speed of sound that is more relevant to phenomenological applications is the one determined at constant entropy per particle, which we are going to compute in this work in \ref{sec:num}. Finally, for completeness, the trace anomaly at finite baryon density is given by
\begin{eqnarray}
         I(T,\mu_{B})&=& \epsilon(T,\mu_{B})-3P(T,\mu_{B})  \\
         &=& Ts(T,\mu_{B})+\mu_{B}\rho_{B}(T,\mu_{B})-4P(T,\mu_{B}).
\nonumber
\end{eqnarray}

\subsection{New numerical procedure}
\label{sec:num}

Now we provide some details on the new numerical approach we developed in the present work, which is crucial to significantly extend the results originally reported in Ref.\ \cite{Critelli:2017oub}. With this new numerical procedure we shall be able to  locate the line of first-order phase transition beyond the critical point of our model and also evaluate several thermodynamic observables across the $(T,\mu_B)$ phase diagram, including the phase transition region, where the numerical computations are particularly complicated to be performed. 
\begin{figure}[h]
     \centering
   \includegraphics[width=0.49\textwidth]{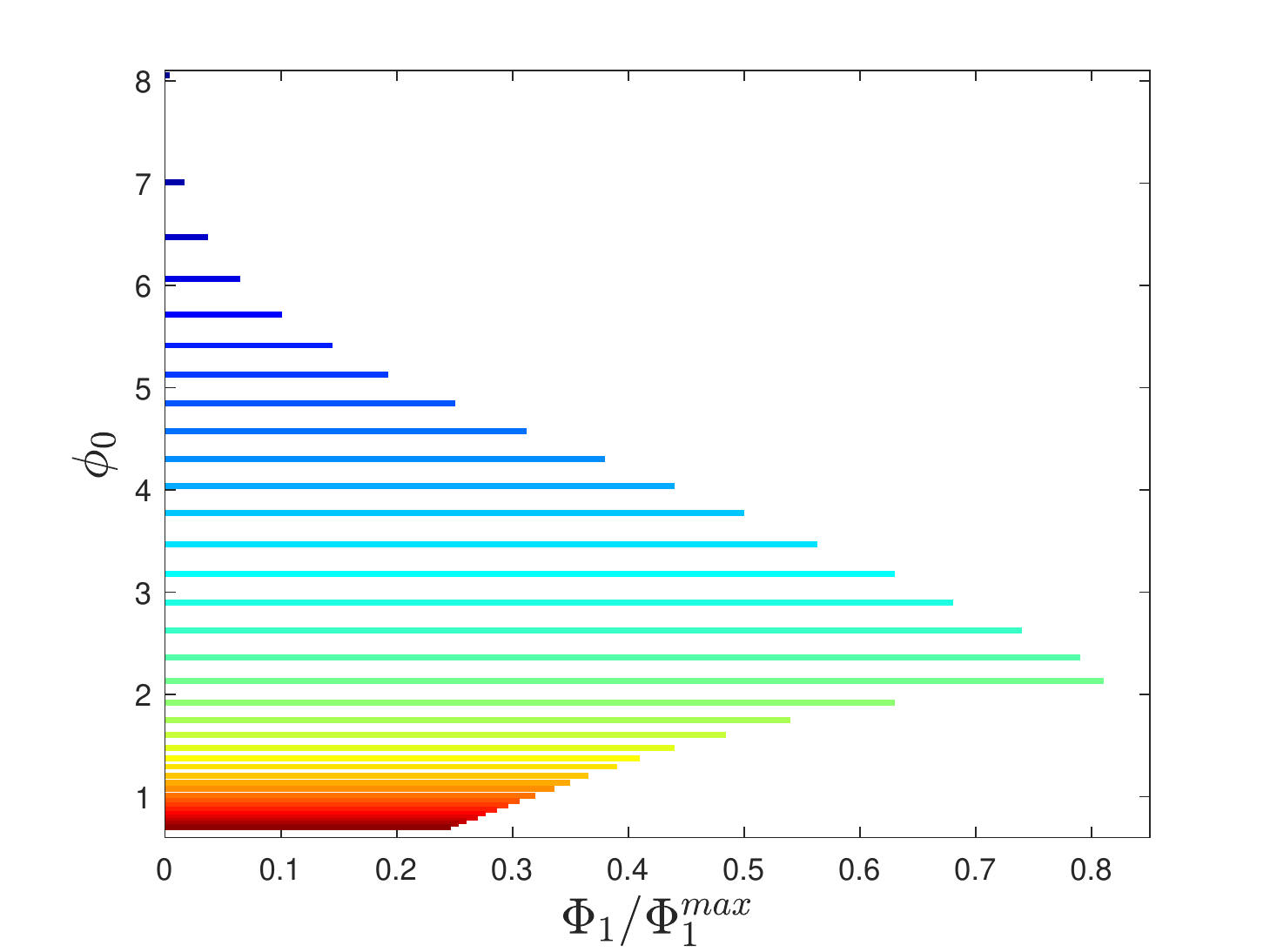}
   \includegraphics[width=0.49\textwidth]{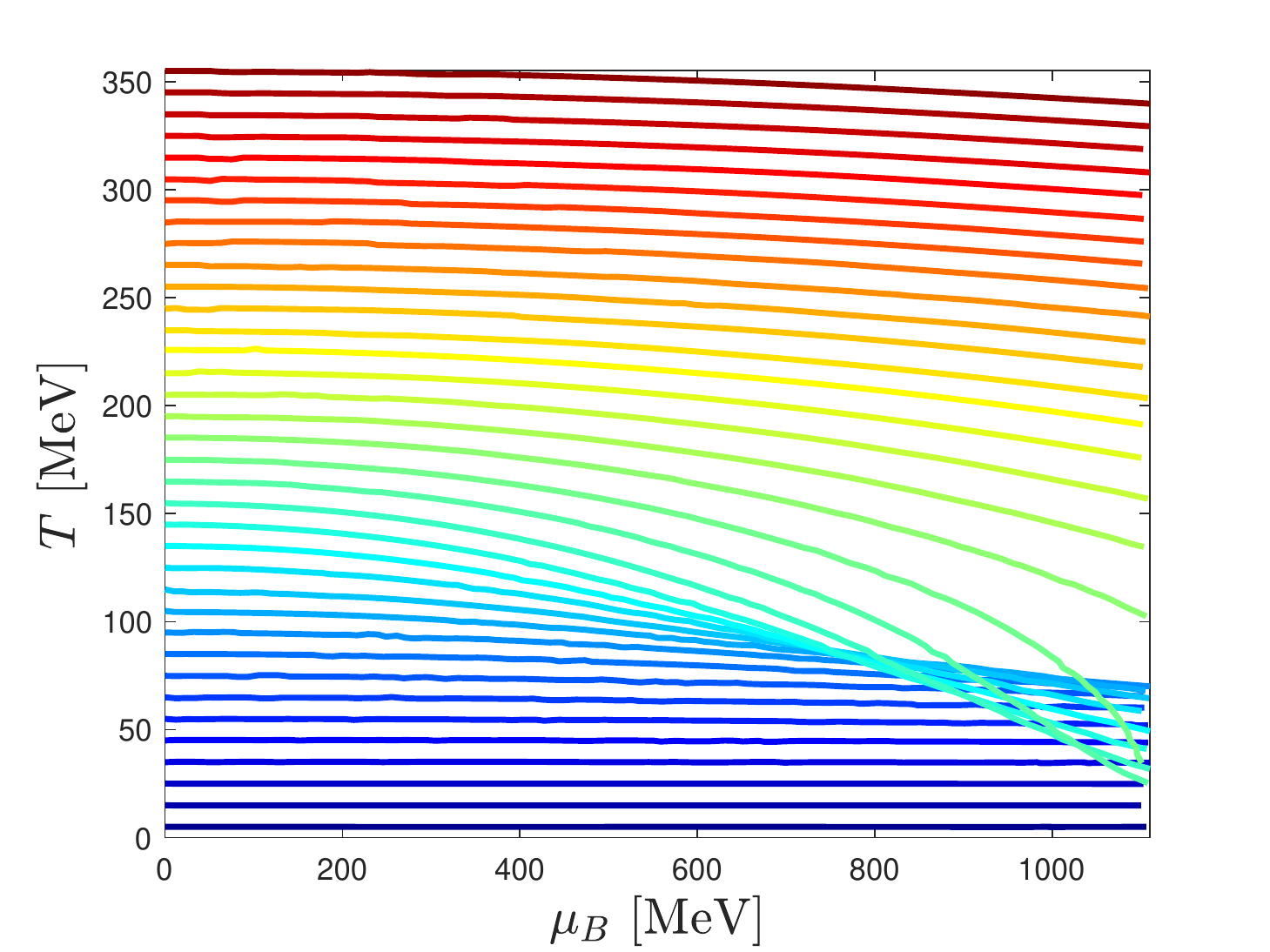}  
    \caption{Example of how the black hole initial conditions should be chosen to map a rectangular region in the QCD phase diagram.}
    \label{fig:mapping2}
\end{figure}
\subsubsection{Integration of the EMD equations of motion}
The equations of motion of the EMD model are solved with the MATLab function "ode113". This function implements a variable-step, variable-order (VSVO) Adams-Bashforth-Moulton PECE solver of order 13. The precision and stability of this method allow us to explore a wider range of black hole boundary initial conditions ($\phi_0, \Phi_1$) than other methods available in MATLab.
The routine used to integrate the EMD fields and find the QCD thermodynamic observables from Eqs. (\ref{eq:T_bh}) --- (\ref{eq:rho_bh}) checks crucial behavior for the stability and physical consistency of the holographic black hole (BH) solutions. A BH-solution is accepted if it satisfies the following requirements:
\begin{itemize}
	\item The integration of the equations of motion \eqref{eq:EoM1} --- \eqref{eq:EoM4} is finite.
	\item The constraint equation \eqref{eq:EoM5} is satisfied.
	\item The dilaton field $\phi(r)$ tends to zero with the correct ultraviolet asymptotics \eqref{UVfields} as we approach the boundary.
	\item The near-boundary behavior of all the other EMD fields also respects the correct ultraviolet asymptotics \eqref{UVfields}.
	\item The metric coefficient $A(r)$ is monotonically increasing.
	\item The Ricci scalar of the black hole background, $R$, is equal to -20 at the ultraviolet radial cutoff $r_{\textrm{max}}$ (meaning that the geometry is already AdS$_5$ at this point).
\end{itemize}

\subsubsection{Mapping QCD thermodynamics from the black hole initial conditions
\label{sec:mapping}}
For the results presented in Ref.\ \cite{Critelli:2017oub}, $2\times 10^{6}$ black holes were generated with initial conditions spanning the rectangle defined by $\phi_{0} \in [0.3,5]$ and $\Phi_{1}\in [0,0.48]\Phi_{1}^{\textrm{max}}(\phi_{0})$. Fig.\ \ref{fig:mapping} shows how an equally spaced, rectangular grid of initial conditions $(\phi_{0},\Phi_{1})$ is mapped into an irregular grid in the $(T,\mu_{B})$ plane generated by the associated black hole solutions.

As seen in Fig.\ \ref{fig:mapping}, a simple rectangular and uniform grid of initial conditions ($\phi_{0}$,$\Phi_{1}$) produces a wide region of the $(T,\mu_B)$ plane which is not covered in the QCD phase diagram (shown in white in the figure). In order to cover the missing section, we introduce here a new way of choosing the black hole initial conditions, which is illustrated in Fig.\ \ref{fig:mapping2}.

We first consider $\Phi_1=0$ (which implies solutions with $\mu_B=0$) and choose the values for $\phi_0$ such that the mapping to the solutions in the temperature axis (at $\mu_B = 0$) is equally spaced in intervals of $0.1$ MeV from $T=2$ MeV to $T=550$ MeV. Next, for each chosen value for $\phi_0$, $\Phi_1$ is varied to map the QCD phase diagram completely up to $\mu_{B}=1100$ MeV, leading to the lines of constant $\phi_0$ shown in Fig.\ \ref{fig:mapping2}. These lines bend in the QCD phase diagram, giving rise to a region with three layers of competing black hole solutions corresponding to the same ($T,\mu_B$) points. In this region, the model is limited at low $T$ by the end of the lines of constant $\mu_B$, were BH-solutions cannot be found using our numerical procedure. It is worth noticing that $\mu_{B}\sim 1100$ MeV is the highest value of $\mu_{B}$ that can be obtained before the BH-solutions for the more curved lines in Fig.\ \ref{fig:mapping2} start diverging and become unstable, which occurs approximately for values of $\Phi_1\gtrsim 0.83\Phi_{1}^{\textrm{max}}(\phi_0)$. In general, a BH-solution cannot be computed when $\Phi_1$ passes this threshold.

For the ensemble of BH-solutions used in the present work, each line of constant $\phi_0$ has 3000 BH-solutions separated uniformly along these lines and corresponding to different values of $\Phi_1$, populating the region of the QCD phase diagram within the rectangle defined by $T \in [2,550]$ MeV and $\mu_B\in [0,1100]$ MeV, without the holes found in Ref.\ \cite{Critelli:2017oub} by using a rectangular grid of initial conditions $(\phi_0,\Phi_1)$, as shown in Fig.\ \ref{fig:mapping}.

\begin{figure}[hbt!] 
    \centering
     \includegraphics[width=0.49\linewidth]{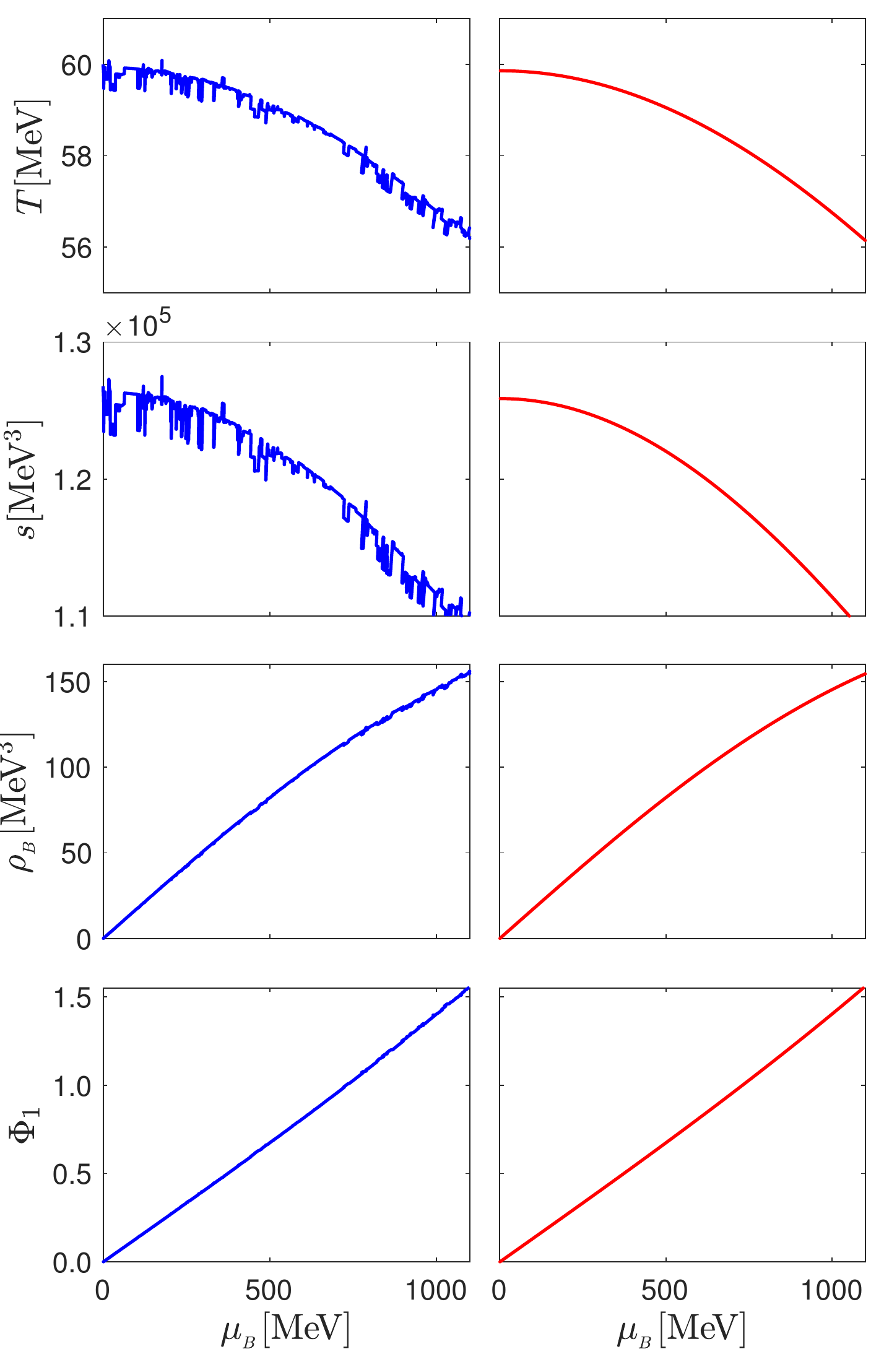}
     \includegraphics[width=0.49\linewidth]{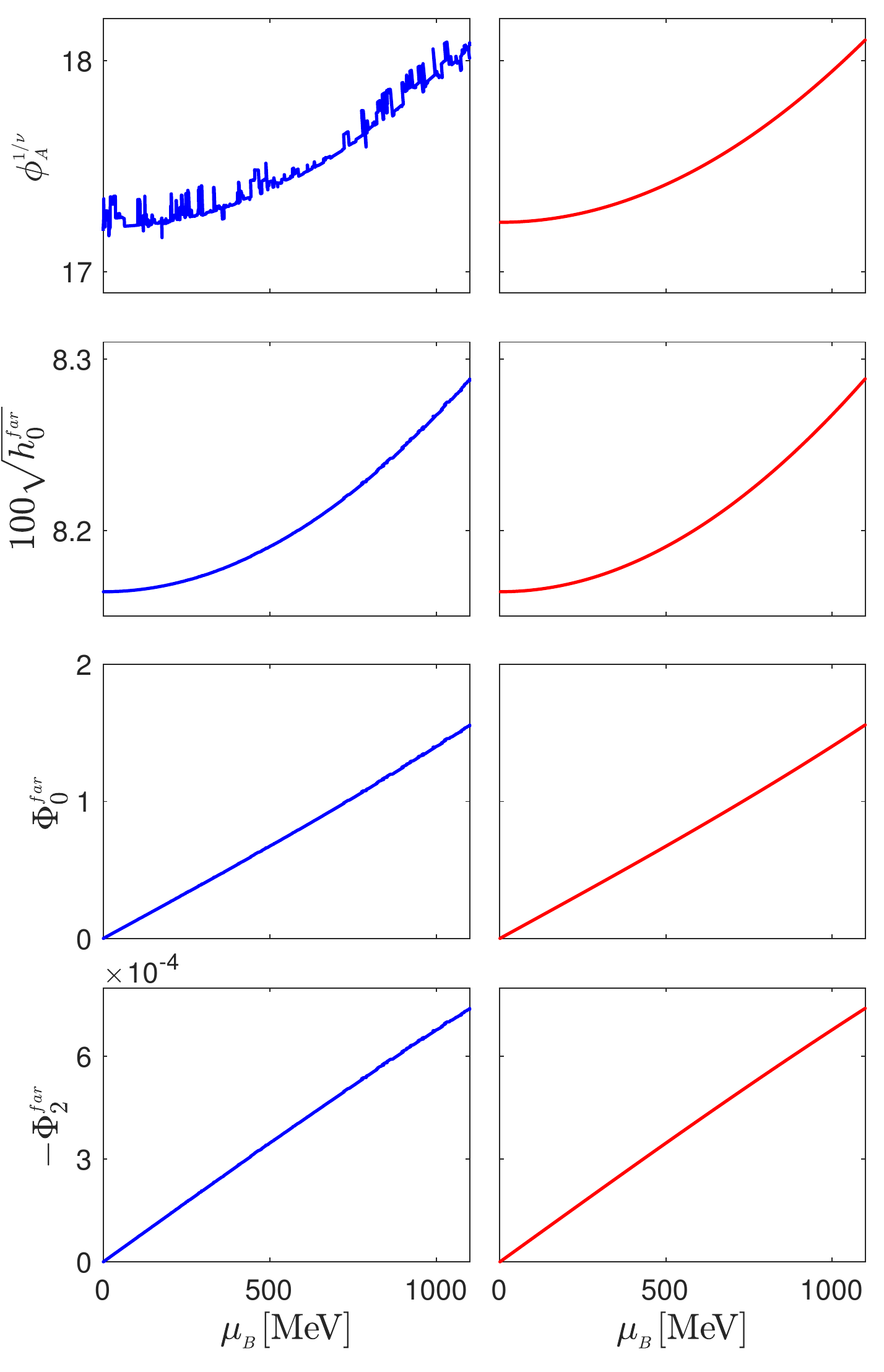}
     \caption{Line of Constant $\phi_0$ for $T=60$ MeV before and after the filtering process.}
    \label{fig:Map60}
\end{figure}

\begin{figure}[hbt!] 
    \centering
     \includegraphics[width=0.49\linewidth]{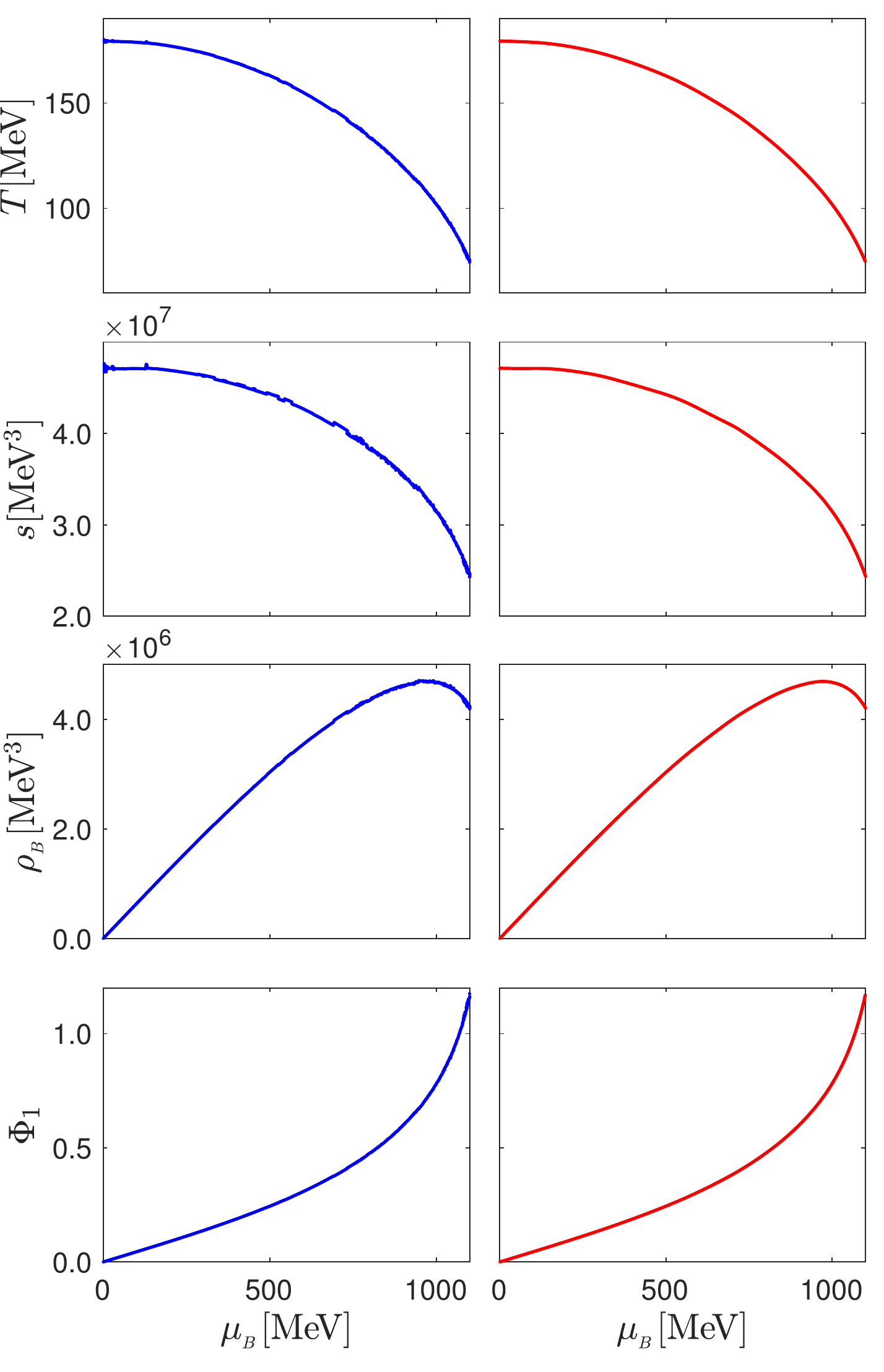}		
     \includegraphics[width=0.49\linewidth]{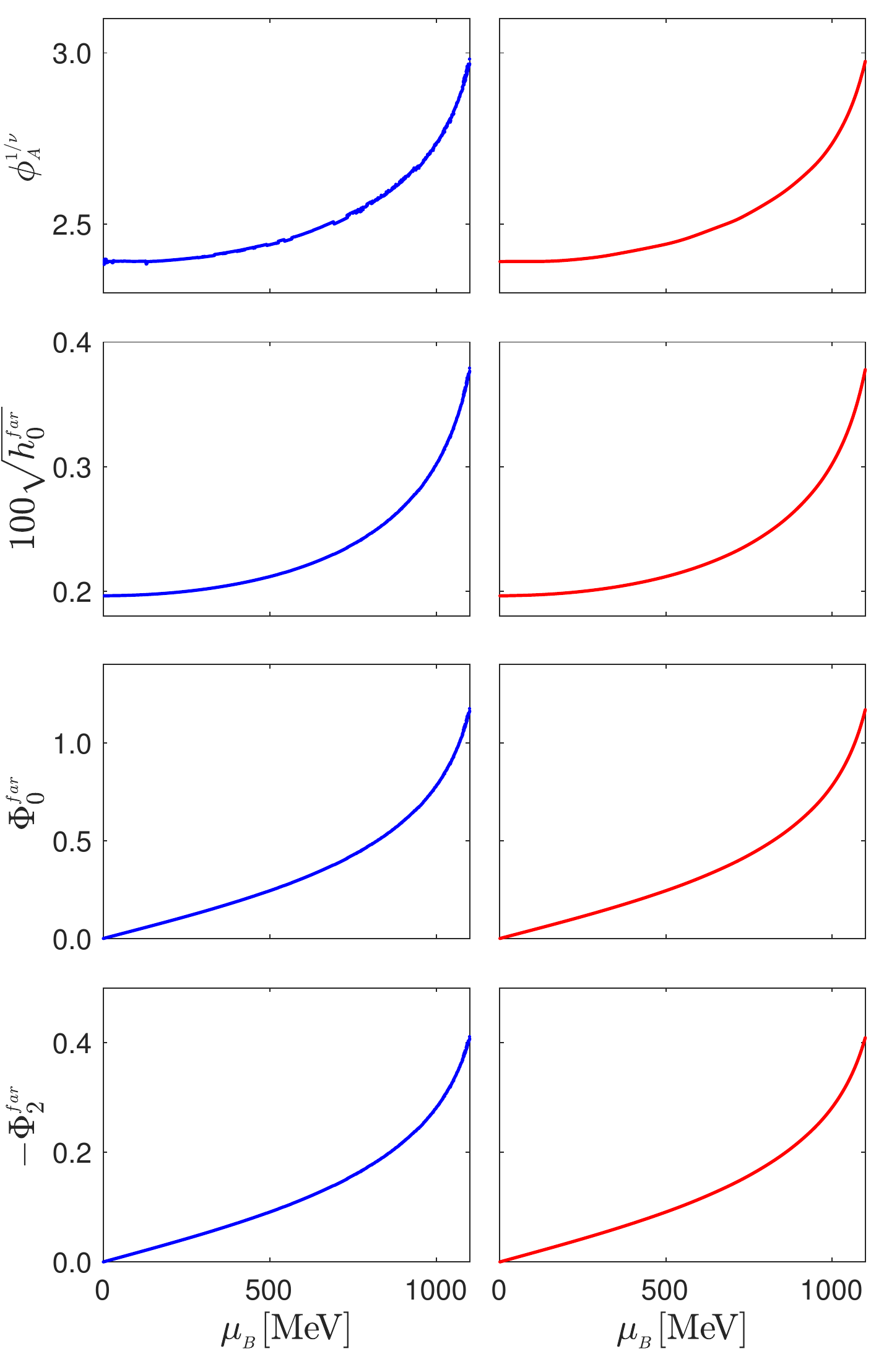}
     \caption{Line of constant $\phi_0$ for $T=180$ MeV before and after the filtering process.}
    \label{fig:Map150}
\end{figure}

\begin{figure}[hbt!] 
    \centering
     \includegraphics[width=0.49\linewidth]{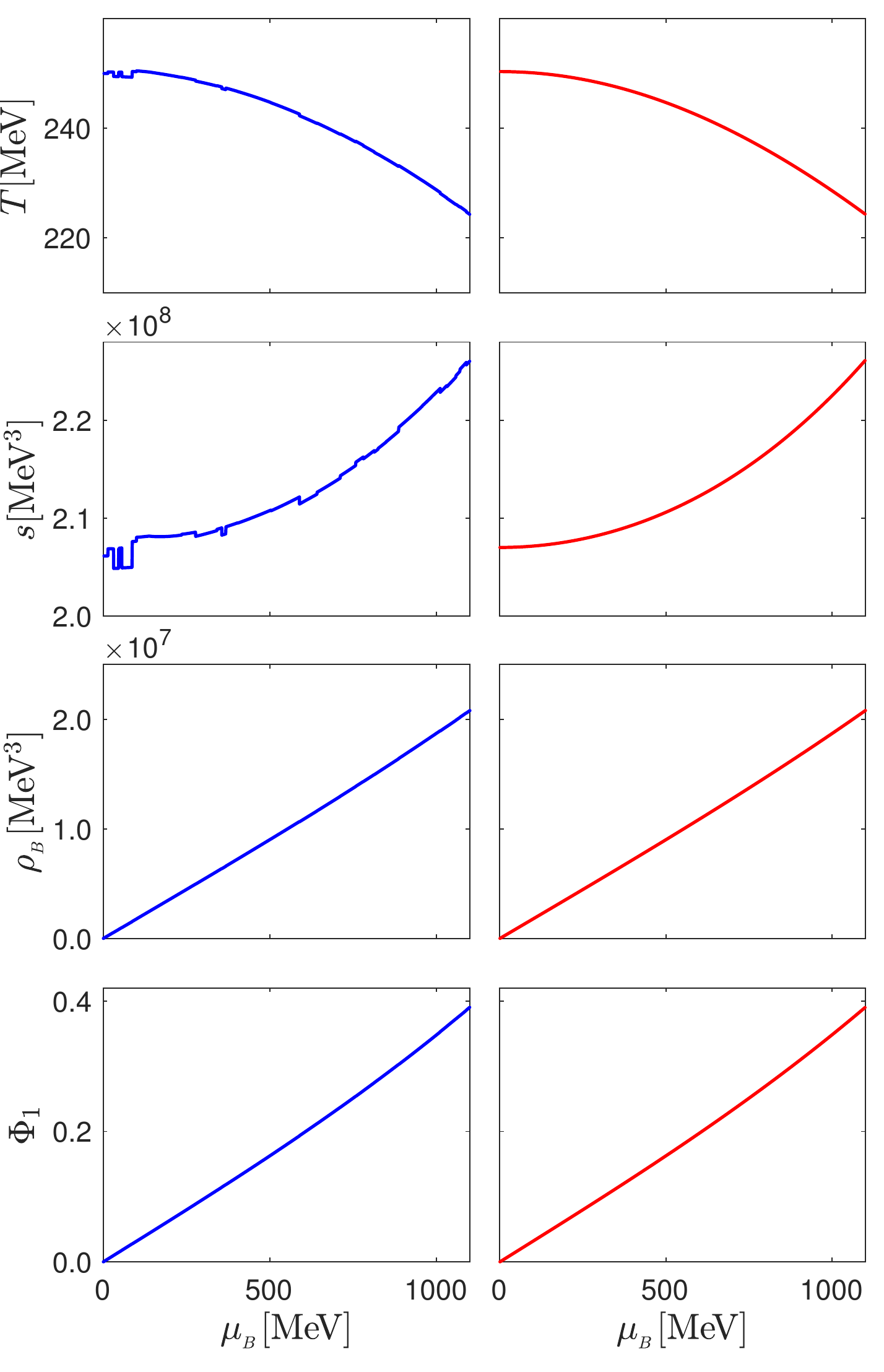}		
     \includegraphics[width=0.49\linewidth]{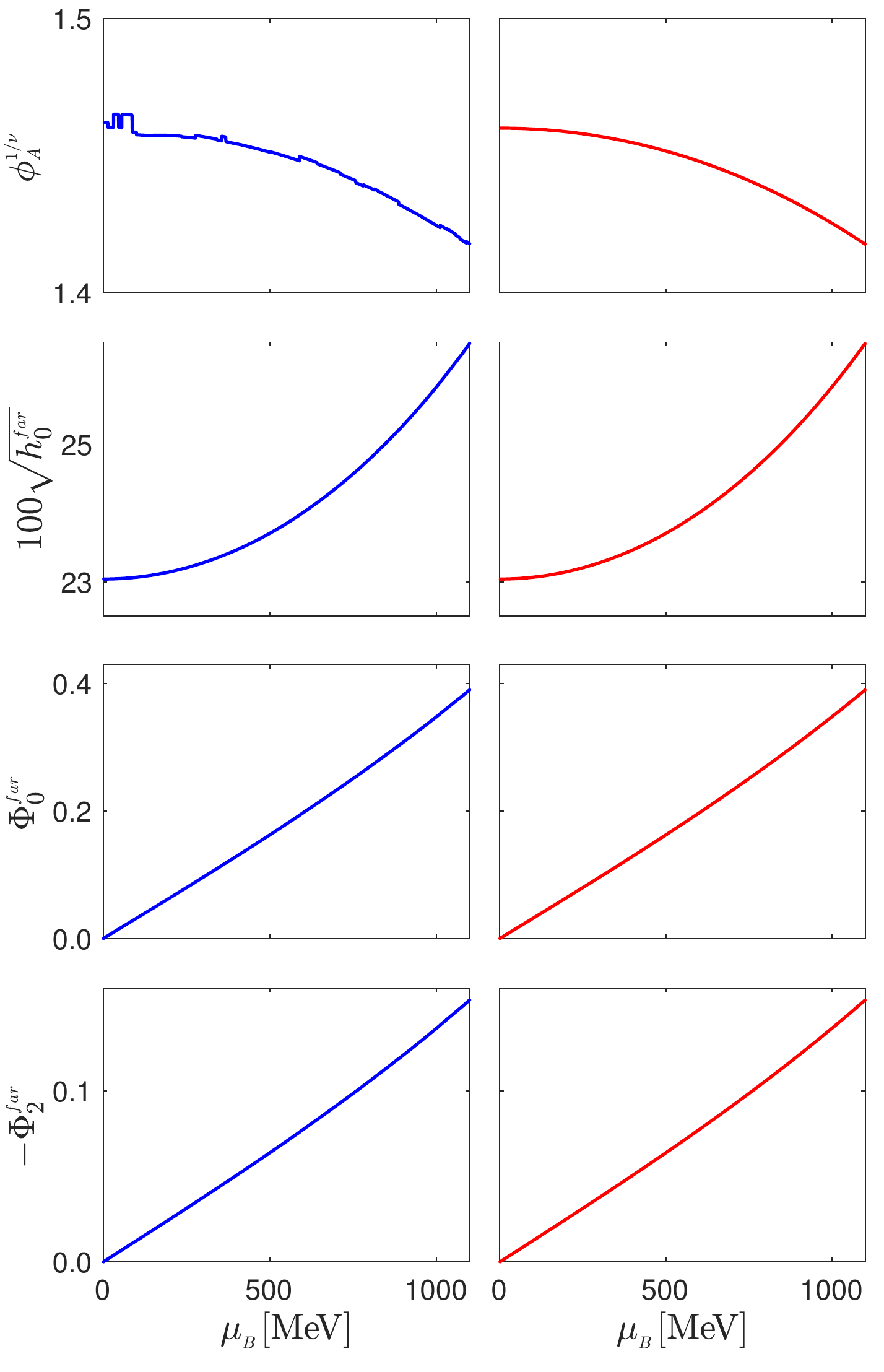}
     \caption{Line of constant $\phi_0$ for $T=250$ MeV before and after the filtering process.}
   \label{fig:Map250}
\end{figure}

The precision of the calculations is significantly affected by numerical noise associated with the fitting of the ultraviolet coefficients in Eq.\ (\ref{UVfields}). The more sensitive coefficient is $\phi_A$, which appears in the holographic thermodynamic formulas (\ref{eq:T_bh}) --- (\ref{eq:rho_bh}) raised to the powers of $-1/\nu$ and $-3/\nu$. The noise associated with the loss of numerical precision is not the same for all lines of constant $\phi_0$, as shown in the left panels of Figs.\ \ref{fig:Map60} --- \ref{fig:Map250}.

The behavior of the ultraviolet coefficients and the thermodynamic variables, as functions of the BH initial conditions, changes for different lines of constant $\phi_0$ as $\Phi_1$ increases. The value of $\Phi_1$ for the lines close to the QCD phase transition (i.e. lines starting between $T=150$ and $T=180$ MeV at $\mu_B=0$) increases much faster than for the other lines and its behavior is not as simple as for the rest of the lines. Therefore, the treatment of the lines is different depending on their location with respect to the QCD transition line.

The strategy to get a smooth mapping is to filter the lines over a large number of BH-solutions. The mapping in Fig.\ \ref{fig:mapping2} contains 3000 BH-solutions per line of constant $\phi_0$. Taking a large number of solutions allows us to treat a noisy line with the appropriate filters without compromising its actual behavior.

For lines with $80 < T < 210$ MeV, $\Phi_1/\Phi_{\textrm{max}}(\phi_0)$ considerably increases (see the color scheme used in Fig.\ \ref{fig:mapping2}, which allows to identify how the different initial conditions map into the $(T,\mu_B)$ plane), and the filtering process consists in smoothing out these lines using a Cubic Smoothing Spline (CSS) filter which only gets rid of big bumps, and then filtering the line with a Savitzky-Golay (SG) filter. SG filters are typically used to smooth out a noisy signal with large noise frequency. For this reason, it is important to prepare the signal with the CSS filter. The SG filter employed during this process uses a polynomial of degree 3 to interpolate each point with its neighbors. The number of neighbors approximate a range of $\pm 20$ MeV.

The rest of the lines are noisy, but the value of $\Phi_1/\Phi_{\textrm{max}}(\phi_0)$ remains small. In this case, the most noisy ultraviolet coefficient is $\phi_{A}^{1/\nu}$, which is corrected by using a simple polynomial fitting of the form $a+bx^2+cx^4$. The remaining ultraviolet coefficients are filtered with the SG filter. Notice that the concavity of $\phi_{A}^{1/\nu}$ changes from positive at small $T$ to negative at large $T$. The region in between is where the BH-solutions can be found with less noise and those lines are the ones that cross the critical point. Figs.\ \ref{fig:Map60} --- \ref{fig:Map250} show the lines of constant $\phi_0$ as functions of $\mu_B$ for different fixed temperatures, before (blue curves) and after (red curves) the filtering process.

\begin{figure} 
    \centering
     \includegraphics[width=0.99\linewidth]{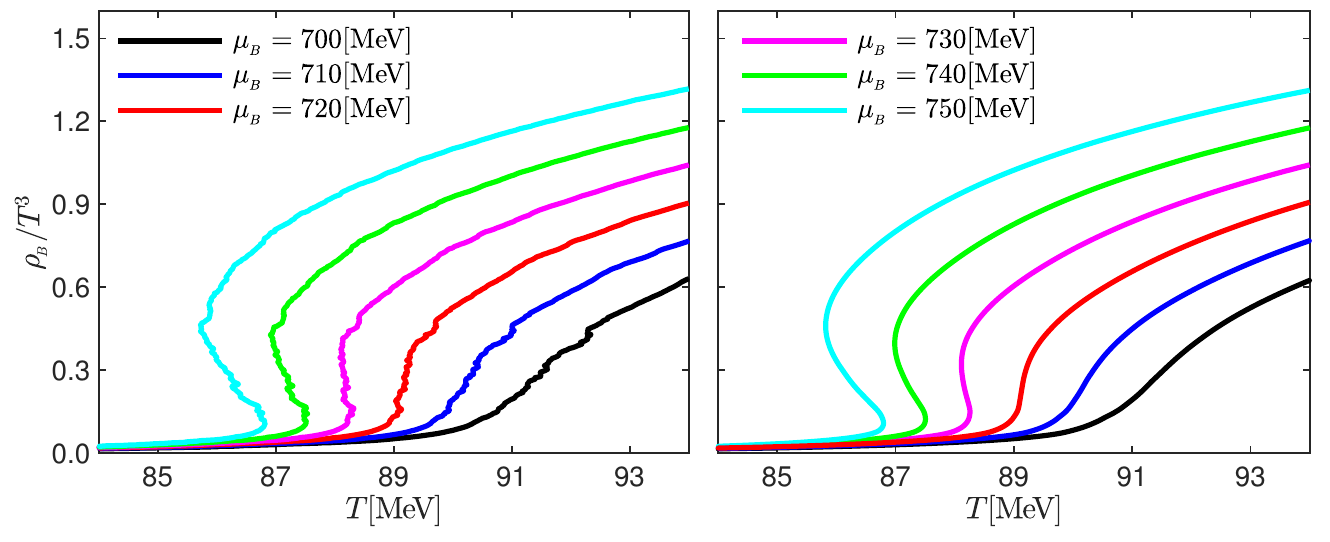}
     \caption{Dependence of the baryon density ($\rho_{B}$) on the temperature $(T)$ for different values of the baryon chemical potential ($\mu_{B}$) before and after the filtering process.}
    \label{fig:rhoB}
\end{figure}

Once the lines of constant $\phi_0$ are corrected, they are fitted with a cubic spline to get lines of constant $\mu_B$. The lines of constant $\mu_B$ are also treated with the SG filter. An example is given in Fig.\ \ref{fig:rhoB}, which shows the baryon density as a function of the temperature for different values of $\mu_B$, before and after the filter.

The lines of constant $\mu_B$ are then fitted with a cubic spline to calculate the pressure, the critical point and the first-order phase transition line. The next step is to calculate lines of constant $T$ which, together with the lines of constant $\mu_B$, are used to take derivatives of the QCD thermodynamic variables.

\subsubsection{Finding the transition line and the QCD critical point}

\begin{figure}[hbt!]
    \centering
    \includegraphics[width=0.5\textwidth]{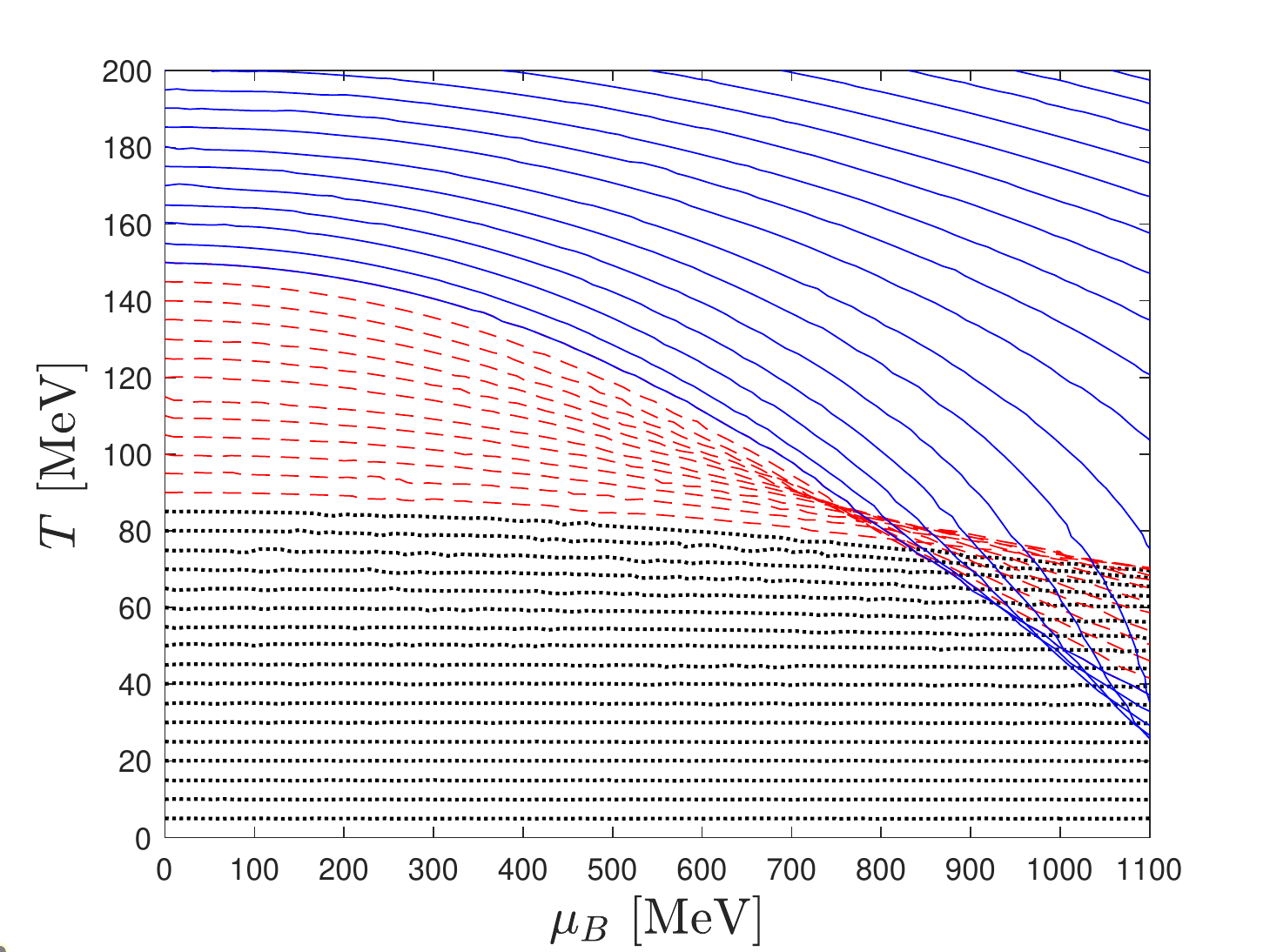}
    \includegraphics[scale=0.5]{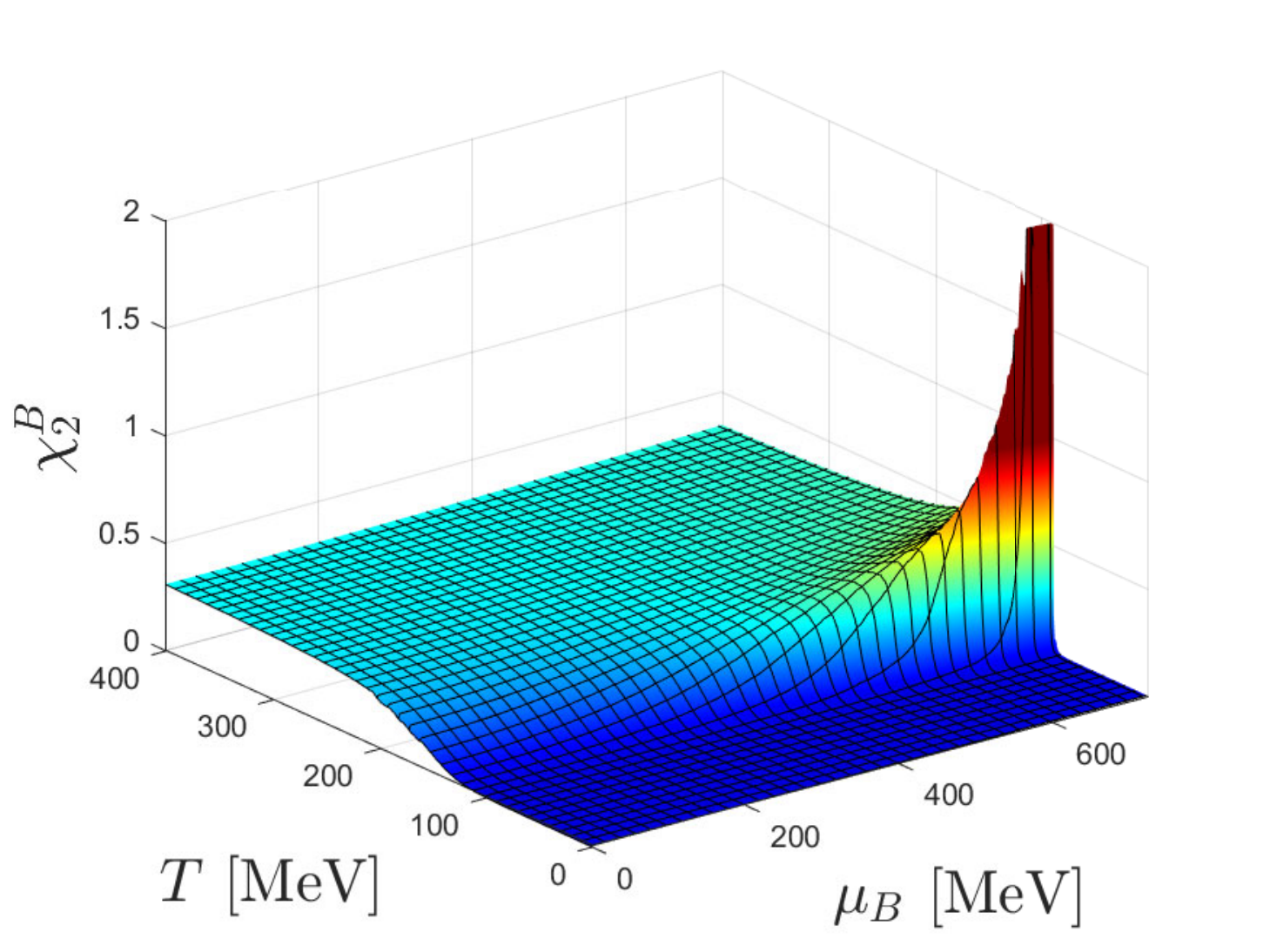}
    \caption{The upper panel shows the mapping from the BH initial conditions ($\phi_{0},\Phi_{1}$) to the QCD phase diagram ($T,\mu_{B}$) before the filtering process. The plot shows how three kinds of lines of constant $\phi_{0}$ are mapped into the ($T,\mu_{B}$) plane, where the crossing of the lines suggest the location of the CEP. The lower panel shows the behavior of the second order baryon susceptibility $\chi_{2}^{B}$ in the $(T,\mu_{B})$ plane. As the chemical potential increases, $\chi_{2}^{B}$ develops a peak that becomes a divergence at the critical point located at $T^{\textrm{CEP}}\sim 89$ MeV and $\mu_{B}^{\textrm{CEP}}\sim 724$ MeV.}
    \label{fig:constantphi}
\end{figure}

The upper panel of Fig.\ \ref{fig:constantphi} shows lines of constant $\phi_{0}$ as $\Phi_{1}$ increases before the filtering process, where we can distinguish three types of lines that define a region of overlapping solutions for the thermodynamics of the holographic EMD model. Three different colors have been used to easily identify the multi-solution region in the figure. The black dotted lines are almost parallel and do not cross each other. Some of the dashed red lines cross each other and also the black lines. Finally, the solid blue lines on the top cross the black and red lines and some cross each other as well. The location where these lines start to intersect can be identified as a candidate point for the critical end point (CEP) in the QCD phase diagram. Due to the presence of the first-order phase transition line, the competing phases may appear as solutions of the equations of motion, although only one minimizes the free energy and represents the true ground state of the system. In the crossover region, one expects only one solution to the equations of motion. However, near the first-order phase transition line, to the right of the critical point, the black hole solutions for the baryon density $\rho_B$ and the entropy density $s$ become multivalued functions of $(T,\mu_B)$. The first-order phase transition line and the multivalued solutions end precisely at the CEP.  

With an equally-spaced rectangular grid in the QCD phase diagram, we can start to analyze the region with multiple solutions. In order to find the exact location of the CEP, one can analyze the second order baryon susceptibility $\chi_{2}^{B}$, which diverges at the critical point. The behavior of $\chi_2^B$ as a function of $T$ and $\mu_B$ is shown in the lower panel of Fig.\ \ref{fig:constantphi}. With this procedure, we find that the critical point is located at $T_{\textrm{CEP}}\sim 89$ MeV and $\mu_{B}^{\textrm{CEP}}\sim 724$ MeV, as originally reported in Ref.\ \cite{Critelli:2017oub}.

\begin{figure*}[hbt!]
    \centering
    \includegraphics[width=0.7\textwidth]{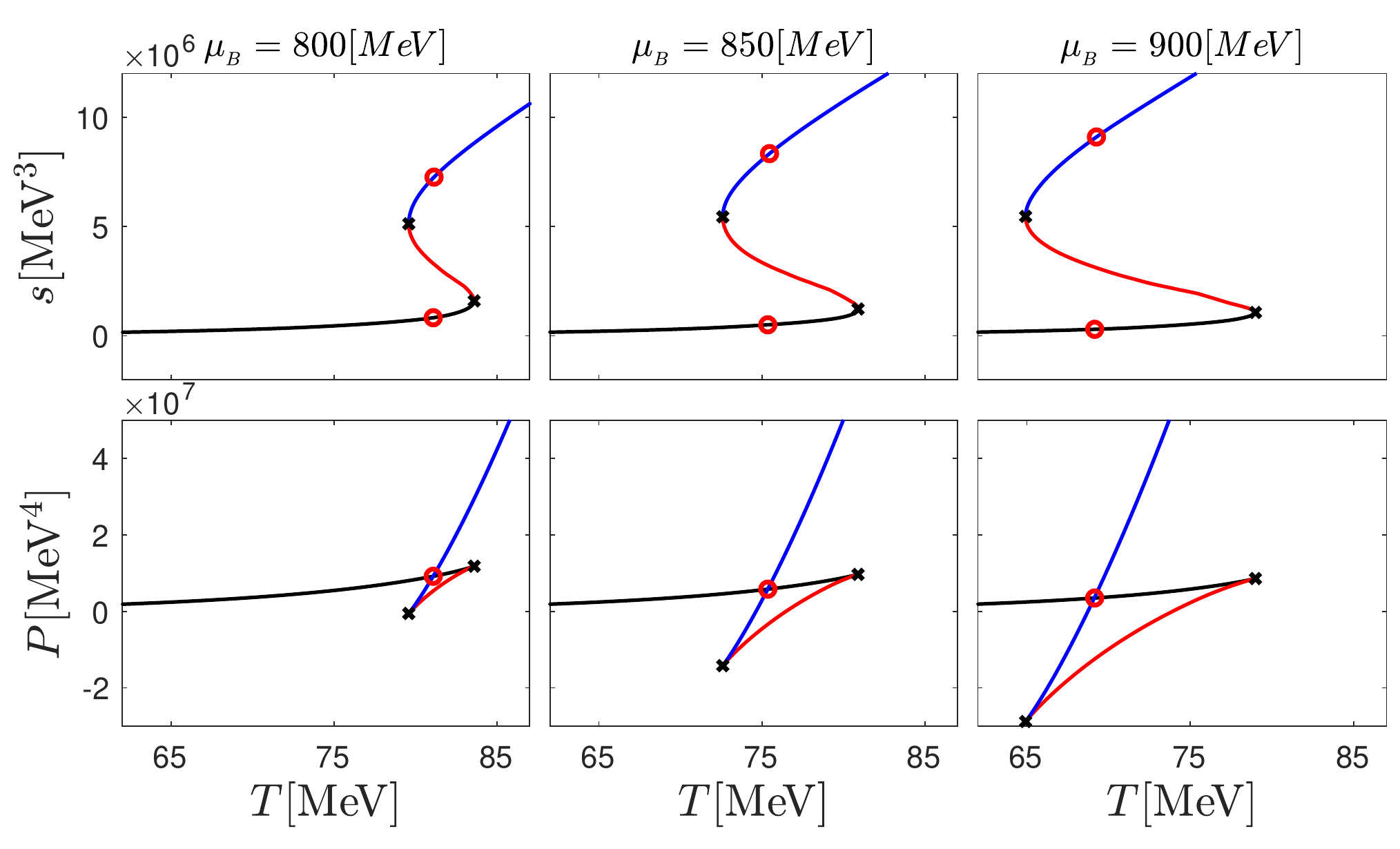}
    \caption{Entropy density $s$ (upper panels) and its integral with respect to the temperature, corresponding to the pressure (lower panels), for three different values of $\mu_{B} > \mu_{B}^{\textrm{CEP}}\sim 724$ MeV.}
    \label{fig:multivalued_entropy}
\end{figure*}

\begin{figure}[hbt!]
    \centering
    \includegraphics[scale=0.5]{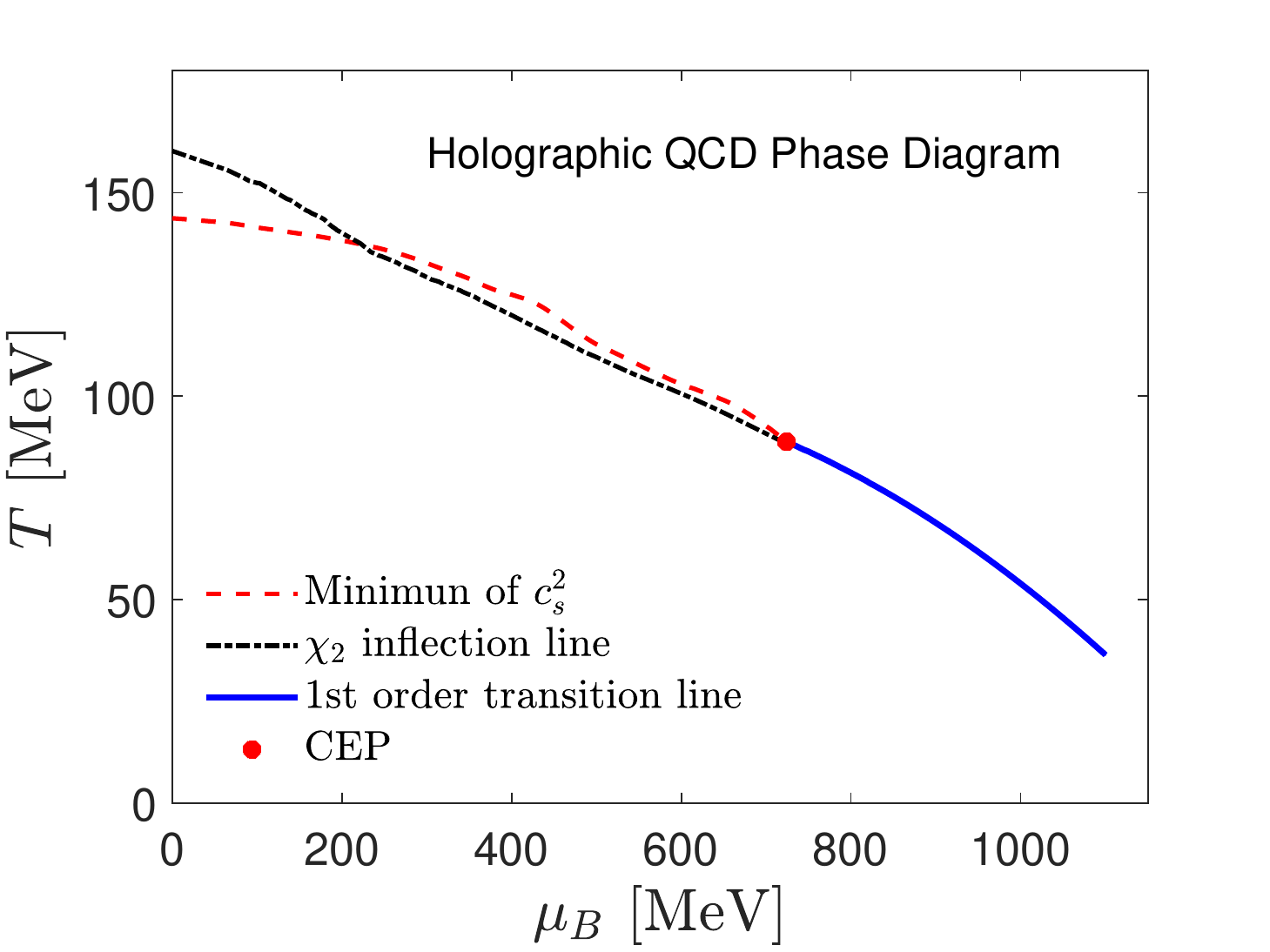}
    
    \caption{The phase diagram of our EMD model. The inflection point of $\chi_2^{B}$ and the minimum of $c_{s}^{2}$ from Eq.\ (\ref{eq:c2s}) are used to characterize the crossover region.}
    \label{fig:transition_line}
\end{figure}

From the highly nonlinear and unequally spaced mapping showed in Fig.\ \ref{fig:mapping}, it is possible to obtain the thermodynamics of QCD on a regular grid in the $(T,\mu_{B})$ plane by means of numerical interpolation as done in \cite{Critelli:2017oub}. In particular, the baryon density $\rho_{B}$ was obtained over a regular grid in the interval $T=[65-450]$ MeV and $\mu_{B}=[0-600]$ MeV via numerical interpolation.

In this work, however, we obtain an equally spaced grid in the $(T,\mu_{B})$ plane directly from the black hole solutions as described in Section \ref{sec:mapping}, by taking the black hole initial conditions as shown in Fig.\ \ref{fig:mapping2}.
One of the advantages of having the thermodynamics over an equally spaced grid in the QCD phase diagram is the opportunity to look at the entropy and baryon density, $s$ and $\rho_{B}$, respectively, over trajectories of constant $T$ or $\mu_{B}$ in the crossover region and near the first-order phase transition line. For instance, for an isotherm at $T>T^{\textrm{CEP}}$ or for slices of constant $\mu_{B}<\mu_{B}^{\textrm{CEP}}$, the entropy density and baryon density are single-valued functions, since they do not cross the first-order phase transition line. On the other hand, for trajectories of constant $T<T^{\textrm{CEP}}$ or $\mu_{B}>\mu_{B}^{\textrm{CEP}}$, i.e. trajectories that cross the first-order phase transition line, $s$ and $\rho_{B}$ become multivalued. Since we are solving the holographic black hole equations of motion, it is reasonable to obtain all extrema of the free energy which corresponds to the coexistence region of not only thermodynamically stable minima, but also thermodynamically metastable and unstable saddle points or maxima. In the top panels of Fig.\ \ref{fig:multivalued_entropy}, we can observe the characteristic multivalued S-shape for the entropy at three different slices of $\mu_{B}>\mu_{B}^{\textrm{CEP}}$, which means that at a given $T$ we have three competing BH-solutions. Precisely at $(T^{\textrm{CEP}},\mu_{B}^{\textrm{CEP}})$, the curves for $s$ and similarly for $\rho_{B}$ cease to be multivalued; this characterizes the end of the first-order phase transition line at the CEP. 

Our approach to characterize the first-order phase transition line was to integrate the entropy with respect to the temperature over the multivalued region, and locate the point where the resulting curve, corresponding to the pressure or to minus the free energy according to Eqs.\ (\ref{eq:Free_energy}) and (\ref{eq:diff_Pressure}), crosses itself. This method is close/analogous to Maxwell's equal area construction, although computationally easier to implement.

It is important to point out that in this work we have only analyzed the thermodynamic observables and identified a line of first order phase transitions ending on a CEP (which was originally predicted for this specific EMD model in Ref.\ \cite{Critelli:2017oub}). However, in principle this phase transition may refer to different aspects of QCD, such as the chiral transition, which in the chiral limit has as an order parameter the chiral condensate, and the deconfinement transition, which in a setup with dynamical quarks has no clear order parameter (since the Polyakov loop is only a legitimate order parameter for the deconfinement transition in the quenched approximation with infinitely heavy quarks). For the present model, we have not calculated either the chiral condensate (this would require considering at least an extra probe action on top of the numerical EMD background solutions), nor the Polyakov loop. Therefore, we cannot specify at this point further details about the nature of this phase transition. Indeed, although for QCD with dynamical quarks at low to moderate values of $\mu_{B}$ the chiral and deconfinement ``transitions'' are a smooth crossover, it is not clear whether those phase transitions are actually located at the same place in the $(T, \mu_{B})$ plane for higher values of $\mu_{B}$.

\begin{figure}[hbt!]
    \centering
    \includegraphics[width=0.48\textwidth]{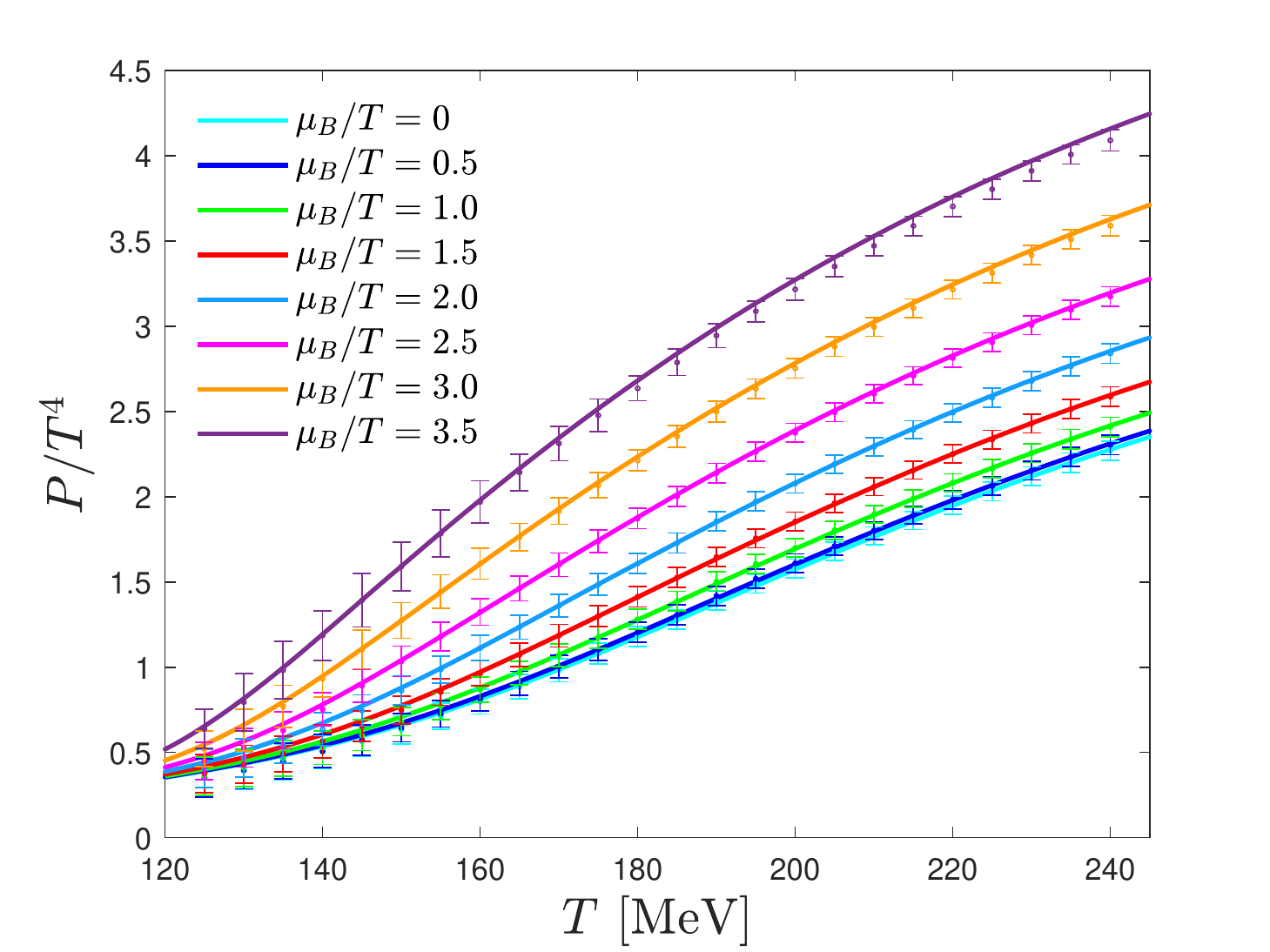}
    \includegraphics[width=0.48\textwidth]{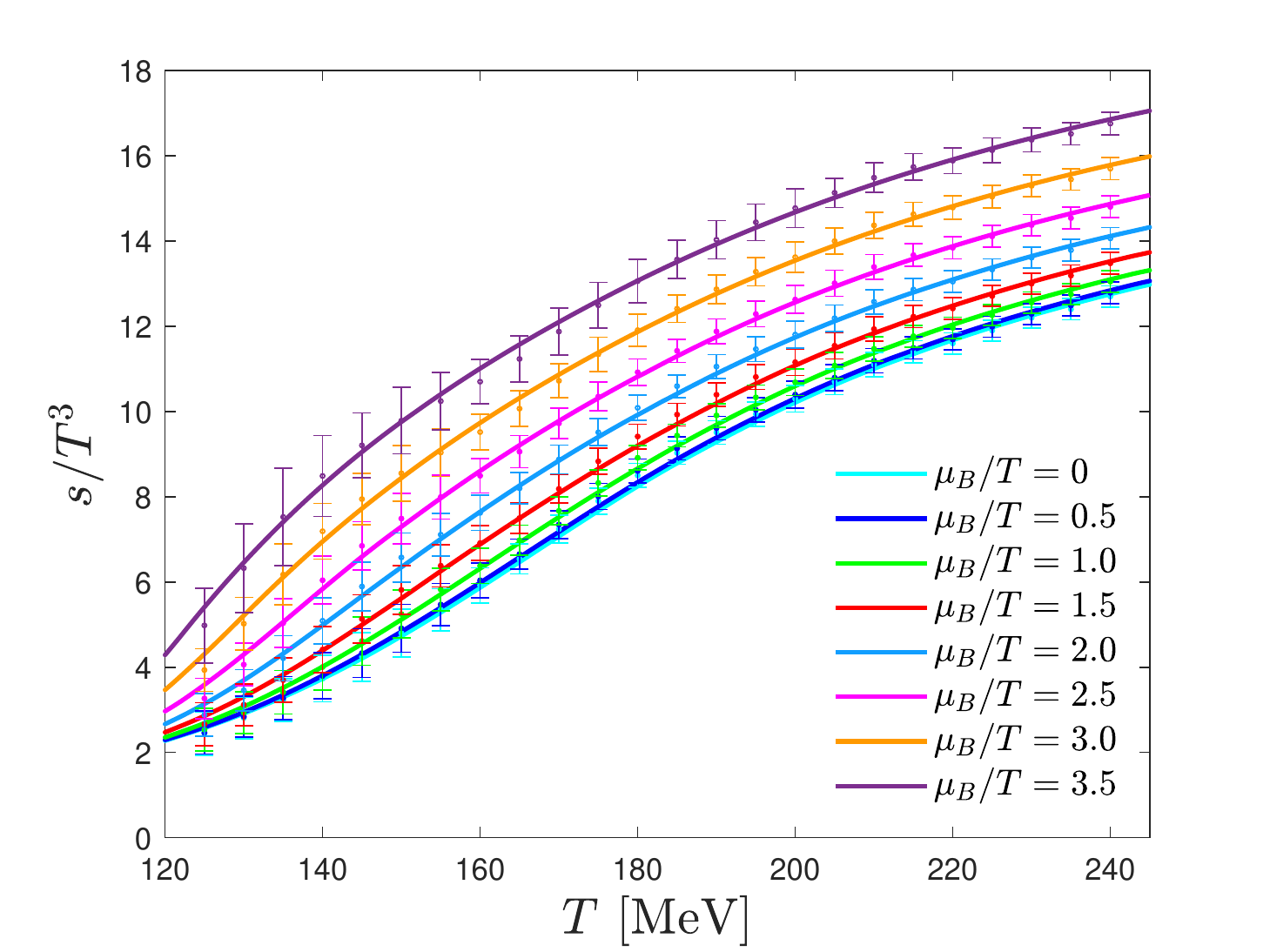}
    \includegraphics[width=0.48\textwidth]{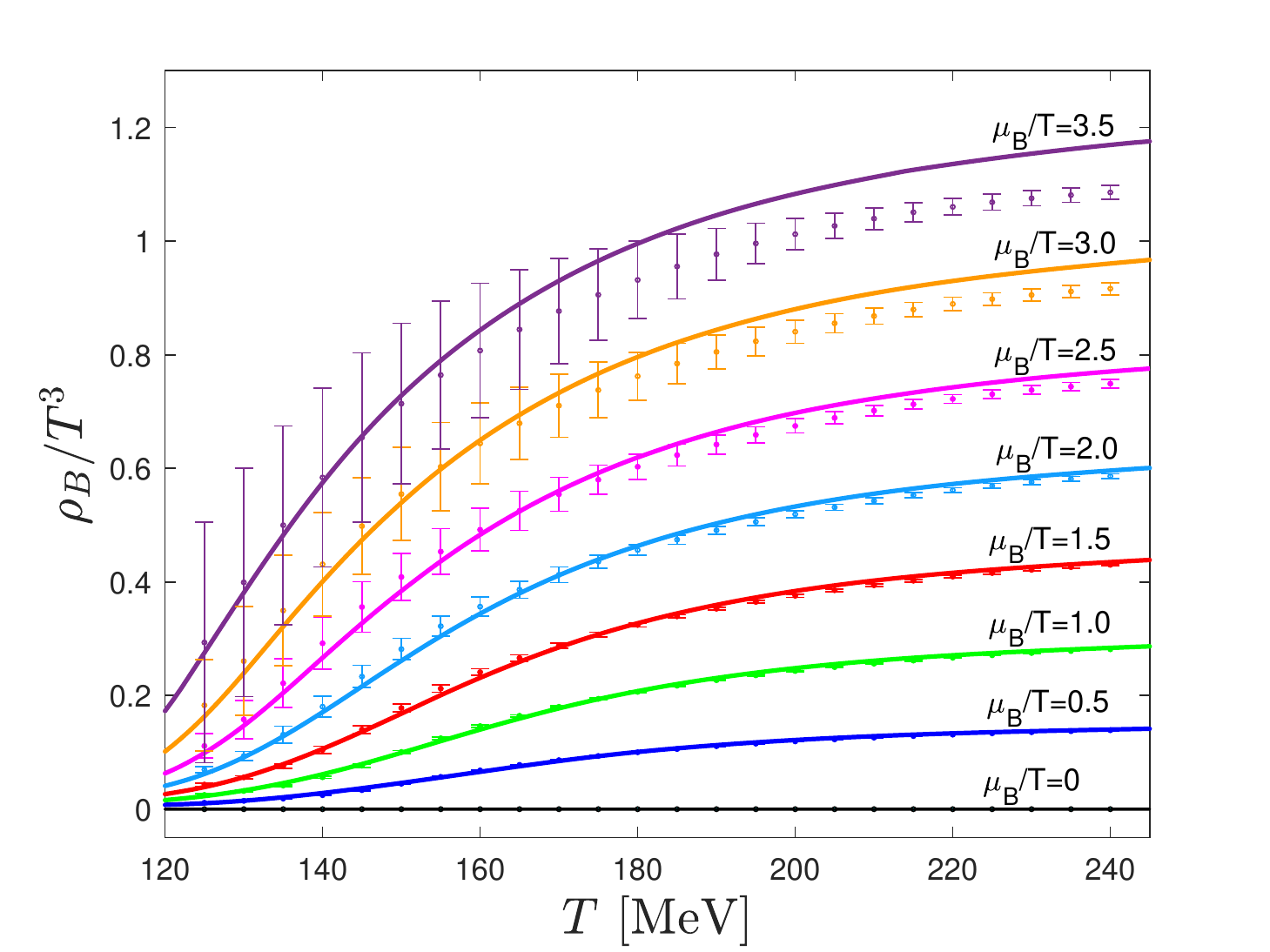}
    
    \caption{Pressure (top panel), entropy density (center panel) and baryon density (bottom panel) as functions of the temperature, for different values of $\mu_B/T$. Our curves are compared to the latest lattice QCD results from Ref.\ \cite{Borsanyi:2021sxv}.}
    \label{fig:lattice_comparison}
\end{figure}

Since the QCD transition from $\mu_B=0$ up to the critical point is a smooth crossover, there is no unique definition of a transition temperature in this region. However, one may try to characterize this quantity as the inflection point or the extrema of observables sensitive to the change of degrees of freedom in the transition between hadrons and a system of quarks and gluons. In fact, as pointed out in Ref.\ \cite{Borsanyi:2010bp}, due to the nature of the crossover and the absence of a real order parameter, several quantities can be used to identify a phase transition in this case, such as the inflection point in the second-order baryon susceptibility or the interaction measure, minimum of the speed of sound and several others. While none of them is a real order parameter, they all exhibit a rapid rise in the vicinity of the transition, and the spread in transition temperature values generated by these different criteria is an indication of the width of the crossover. In this work, we have chosen to characterize the transition in the crossover region by both the inflection point of the second order baryon susceptibility $\chi_2^{B}$, and the minimum of the square of the speed of sound $c_{s}^{2}$ at constant entropy per particle. While the corresponding transition temperatures are not the same in the crossover region, they do come together at the critical point, as shown in Fig. \ref{fig:transition_line}.

The baryon susceptibilities are generally defined as:
\begin{equation}\label{eq:susceptibilities}
    \chi_{n}=\frac{\partial^n (P/T^{4})}{\partial(\mu_{B}/T)^{n}},
\end{equation}
which are basically the coefficients in the Taylor expansion of the pressure
\begin{equation}\label{eq:pressure_tayor}
    \frac{P(T,\mu_{B})-P(T,\mu_{B}=0)}{T^{4}}=\sum_{n=1}^{\infty}\frac{1}{(2n)!}\chi_{2n}(T)\left(\frac{\mu_{B}}{T}\right)^{2n},
\end{equation}
and the baryon density
\begin{equation}\label{eq:density_taylor}
    \frac{\rho_{B}(T,\mu_{B})}{T^{3}}=\sum_{n=1}^{\infty}\frac{1}{(2n-1)!}\chi_{2n-1}(T)\left(\frac{\mu_{B}}{T}\right)^{2n-1}.
\end{equation}
In particular, $\chi_2^B$ measures the equilibrium response of the baryon density to a change in the chemical potential of the medium. The square of the speed of sound at constant entropy per particle is defined as $c^{2}_{s}=\left(\partial P/\partial\epsilon\right)_{s/\rho_{B}}$, but this definition is not practical when one wants to calculate it on top of a $(T,\mu_{B})$ grid of points. For this reason, it is advantageous to rewrite this state variable in terms of derivatives of the pressure along lines of constant temperature or chemical potential only \cite{Parotto:2018pwx,Floerchinger:2015efa}:

\begin{equation}\label{eq:c2s}
c_{s}^{2}=\frac{\rho_{B}^{2}\partial_{T}^{2}P-2s\rho_{B}\partial_{T}\partial_{\mu_{B}}P +s^{2}\partial_{\mu_{B}}^{2}P}{(\epsilon+P)[\partial_{T}^{2}P\partial_{\mu_{B}}^{2}P-(\partial_{T}\partial_{\mu_{B}}P)^{2}]}.
\end{equation}
The lines in the phase diagram corresponding to the minimum of the square of speed of sound computed from Eq.\ (\ref{eq:c2s}) (red, dashed) and to the inflection point of $\chi_{2}^{B}$ (black, dash-dotted) are shown in Fig.\ \ref{fig:transition_line}. As shown in the figure, the two lines are separated in temperature at $\mu_B=0$, while they meet at the critical point. The first-order phase transition line, computed using the scheme implemented in this paper, is plotted as a blue full line in Fig.\ \ref{fig:transition_line}.

The dependence of the transition temperature (defined by the minimum of $c^{2}_{s}$) on the chemical potential in the crossover region can be characterized by the following truncated series 
\begin{equation} \label{eq:crossover_param}
\frac{T_{c}(\mu_{B})}{T_{c}(0)}=1-\kappa_{2}\left(\frac{\mu_{B}}{T_{c}(0)}\right)^{2}- \kappa_{4}\left(\frac{\mu_{B}}{T_{c}(0)}\right)^{4},
\end{equation}
with $T_{c}(0)=143.4$ MeV, $\kappa_{2}=0.0187$, and $\kappa_{4}=-0.00158$. In the case of the most recent lattice QCD transition line obtained from the inflection point of the chiral condensate and its susceptibility in the crossover region, the values of $\kappa_{2}$ and $\kappa_{4}$ are $0.0153(18)$ and $0.00032(67)$, respectively \cite{Borsanyi:2020fev}. It should be noted, however, that these expansion coefficients for the minimum of $c_s^2$ and for the inflection of the chiral condensate do not need to agree, since the corresponding transition curves are actually different in the crossover region.

\subsection{Equation of state}
\label{sec:EOS}

\begin{figure*}[htp!]
\centering
   \includegraphics[width=0.48\textwidth]{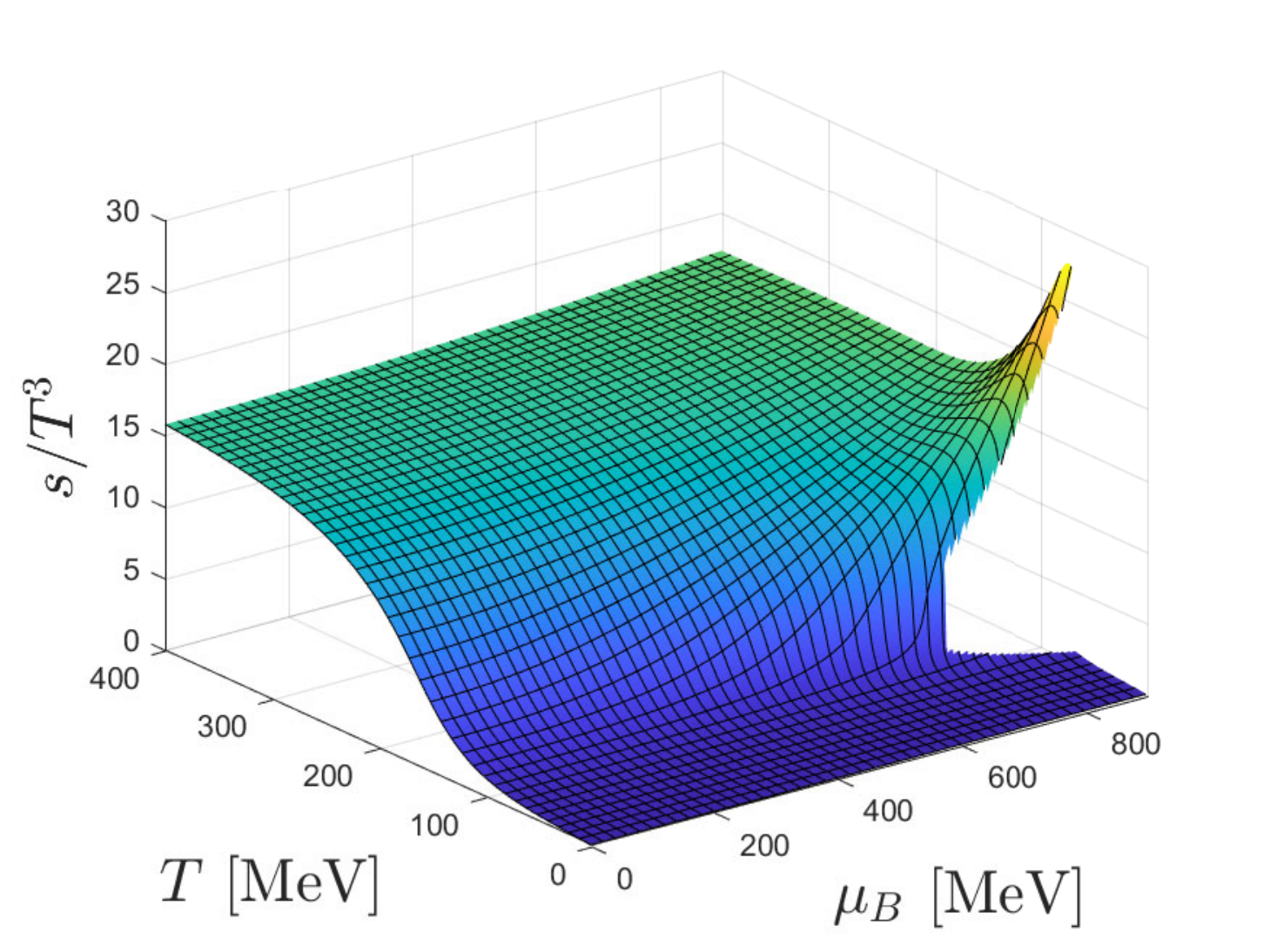}
   \includegraphics[width=0.48\textwidth]{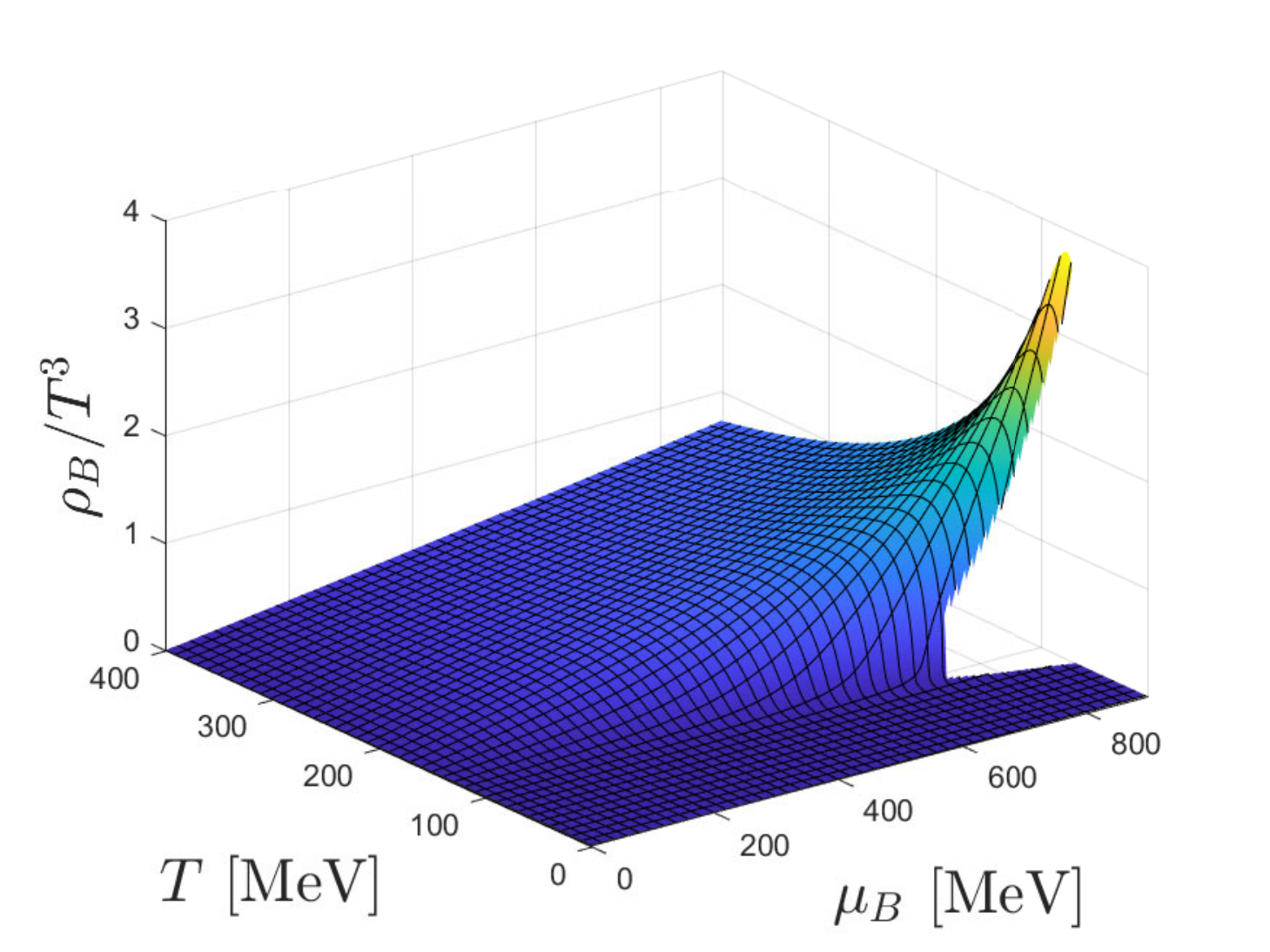}  
   \newline
\centering
  \includegraphics[width=0.48\textwidth]{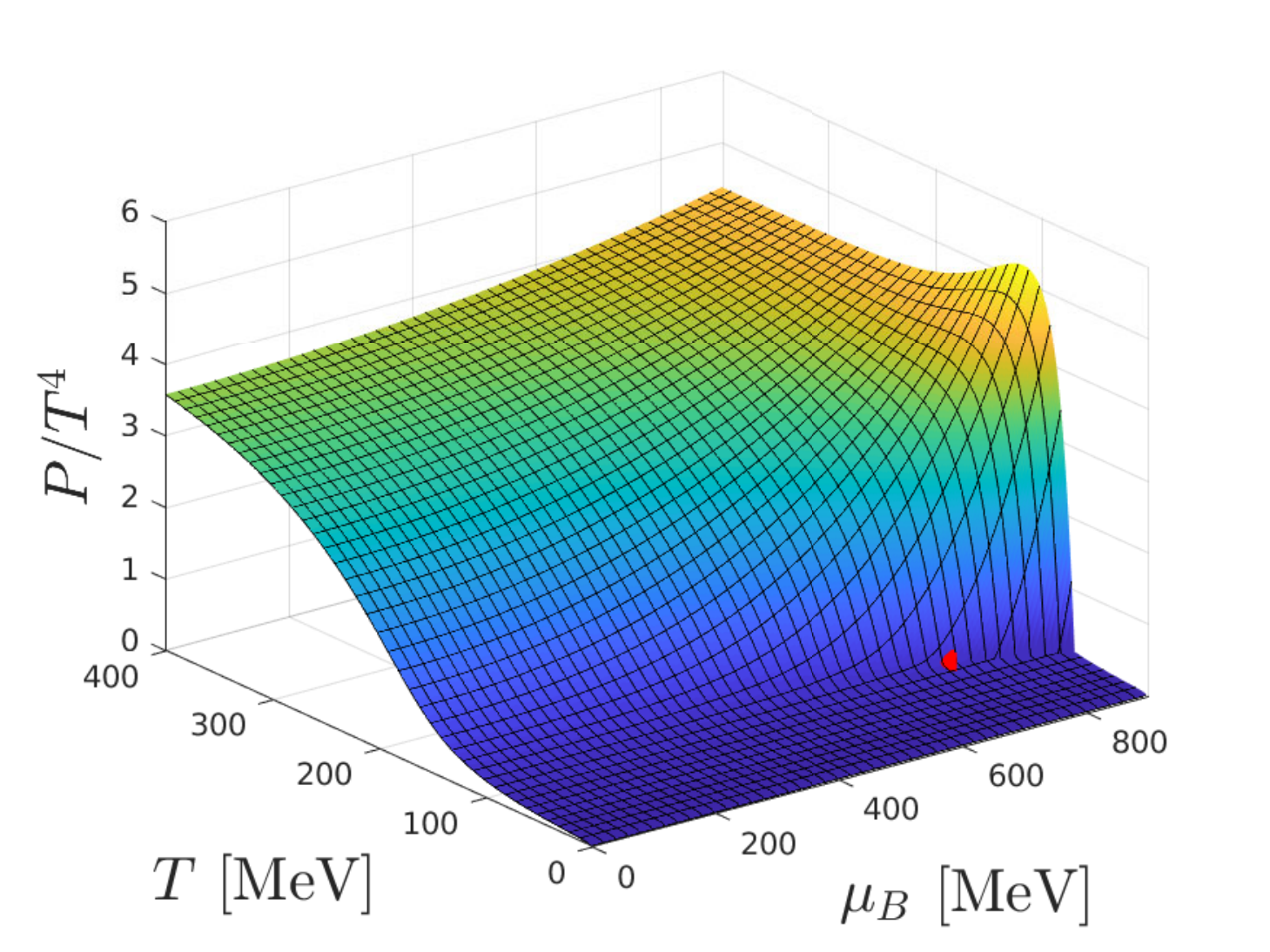}
  \includegraphics[width=0.48\textwidth]{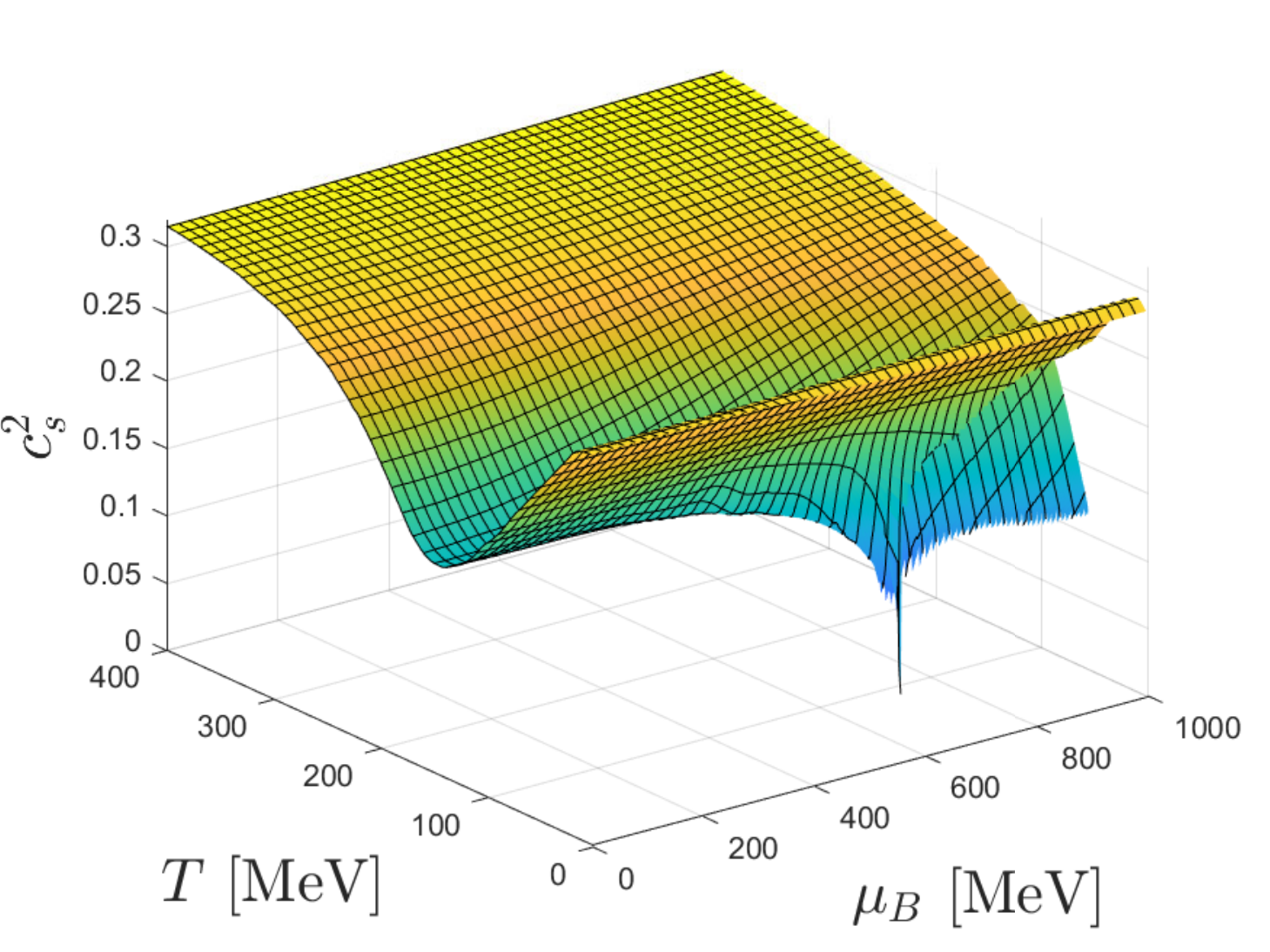}
    \caption{Entropy density (top left), baryon density (top right), pressure (bottom left), and square of the speed of sound at constant entropy over baryon number as computed via Eq.\ (\ref{eq:c2s}) (bottom right). The CEP is shown on the pressure surface as a red dot.}
    \label{fig:full_EoS}
\end{figure*}

The comparison between the holographic EMD equation of state and the Taylor-expanded lattice QCD equation of state up to $\mu_B/T\le 2$ \cite{Bazavov:2017dus,Gunther:2016vcp} was presented in Ref.\ \cite{Critelli:2017oub}. In that work, we also predicted the location of the QCD CEP to lie at $(T,\mu_B)\sim(89,724)$ MeV but, at that time, due to numerical difficulties, we were unable to identify the location of first-order phase transition line beyond the CEP and also to calculate the thermodynamic observables in the phase transition region. These numerical difficulties were solved in the present work through the new developments discussed in previous sections, namely: i) the new way of choosing the BH initial conditions illustrated in Fig.\ \ref{fig:mapping2}, which allowed us to cover a much larger region of the $(T,\mu_B)$ phase diagram, including the location of the first-order phase transition line; ii) the filtering scheme, which allowed us to obtain smooth results for the physical observables in the phase transition region, where their computation is plagued by strong numerical noise. In fact, without the filtering process, the result for the entropy density in the phase transition region is so noisy that it becomes impossible to obtain sensible results for the pressure by integrating the entropy, which in turn makes it impossible to correctly identify the thermodynamically stable BH-solutions in the multi-solution phase transition region and the first-order phase transition line.

With the aforementioned technical developments, in the present work we largely extend the coverage of the EMD model on the $(T,\mu_B)$ plane and present our results for the equation of state in a broader region of the phase diagram and also compare our results with the most up-to-date lattice data for the QCD equation of state, which is now available up to the unprecedentedly high value of $\mu_B/T=3.5$ \cite{Borsanyi:2021sxv}.

In Fig.\ \ref{fig:lattice_comparison} we show how the holographic EMD equation of state compares to the lattice data from Ref.\ \cite{Borsanyi:2021sxv}. We see that the entropy density predicted by the EMD model is in quantitative agreement with the lattice results for all the values of $T$ and $\mu_B$ currently covered by lattice simulations. Regarding the pressure, there is also quantitative agreement for most of the values of $T$ and $\mu_B$, although the EMD result starts to deviate from the lattice outcome for the pressure for $\mu_B/T\ge 3.5$ in the high temperature region with $T\gtrsim 220$ MeV. With respect to the baryon density, our results are in quantitative agreement with the lattice simulations for all the values of $\mu_B/T$ and temperatures up to $T\sim 190$ MeV, although the holographic EMD prediction overestimates the lattice results for the baryon density at high temperatures $T\gtrsim 190$ MeV when $\mu_B/T\gtrsim 2.5$. Interestingly enough, for lower temperatures $T\lesssim 190$ MeV, where the transition from a hadron gas to the quark-gluon plasma phase takes place, the holographic EMD predictions for the entropy density, the pressure, and the baryon density are in quantitative agreement with the lattice results all the way up to $\mu_B/T = 3.5$, which suggests that our prediction for the behavior of the QCD phase transition at nonzero baryon chemical potentials is robust.

In Fig.\ \ref{fig:full_EoS} we show the surface plots of the entropy density, baryon density, pressure and square of the speed of sound in the $(T,\mu_{B})$ plane. We obtained the temperature, baryon chemical potential, entropy and baryon density directly from the holographic dictionary given by Eqs.\ (\ref{eq:T_bh}) --- (\ref{eq:rho_bh}). The pressure is found by integrating the entropy with respect to the temperature at constant baryon chemical potential as in Eq.\ (\ref{eq:diff_Pressure}), and it can also be computed as the integral of the baryon density with respect to the baryon chemical potential along isotherms as suggested in Eq.\ (\ref{eq:dF}), which produced the same result and served as a cross check. The second order baryon susceptibility $\chi_2^B$ shown in Fig.\ \ref{fig:constantphi} is found as the derivative of the baryon density along the chemical potential direction.

\begin{figure*}[htp!]
    \centering
    \includegraphics[width=0.49\textwidth]{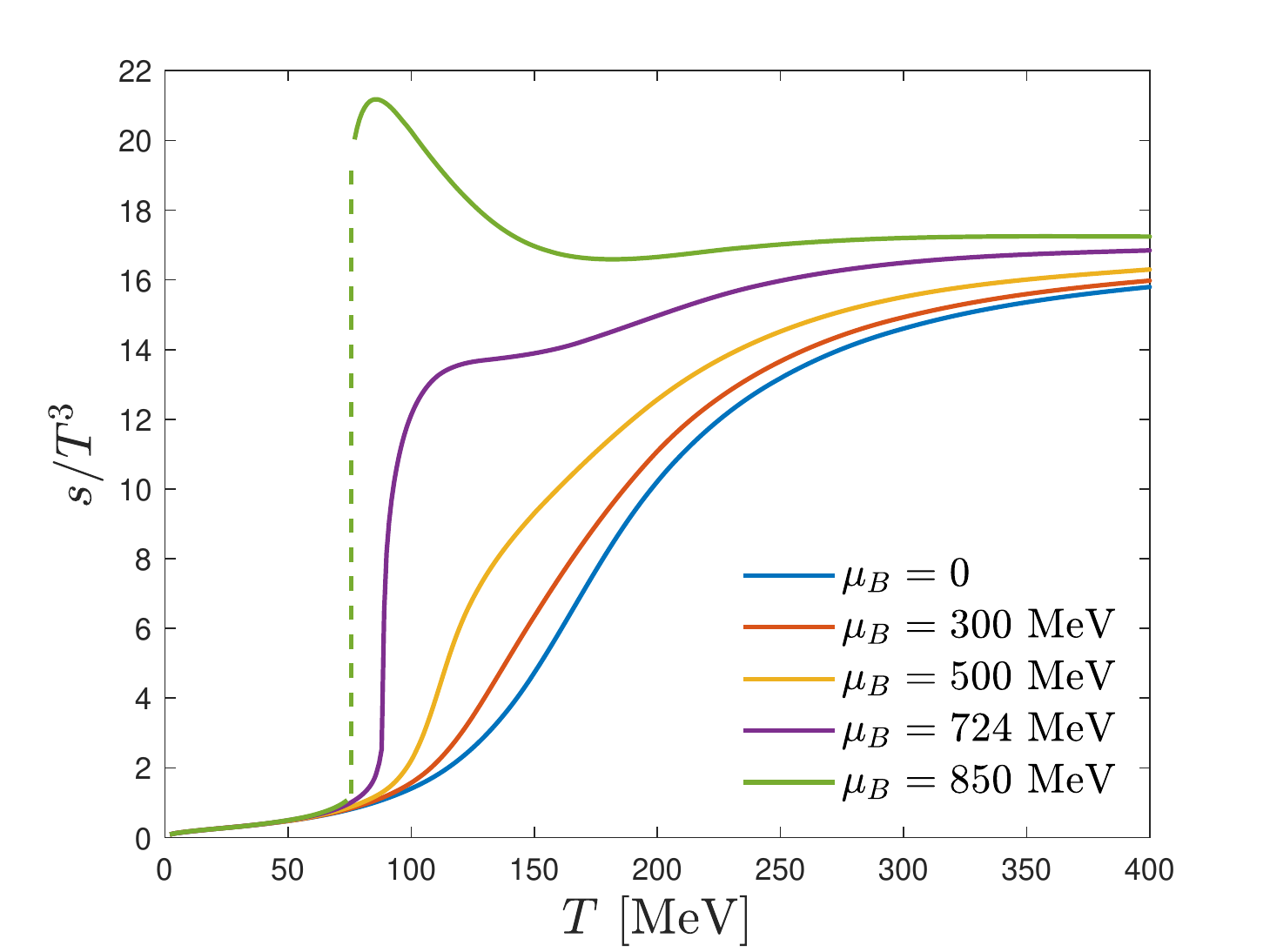}
    \includegraphics[width=0.49\textwidth]{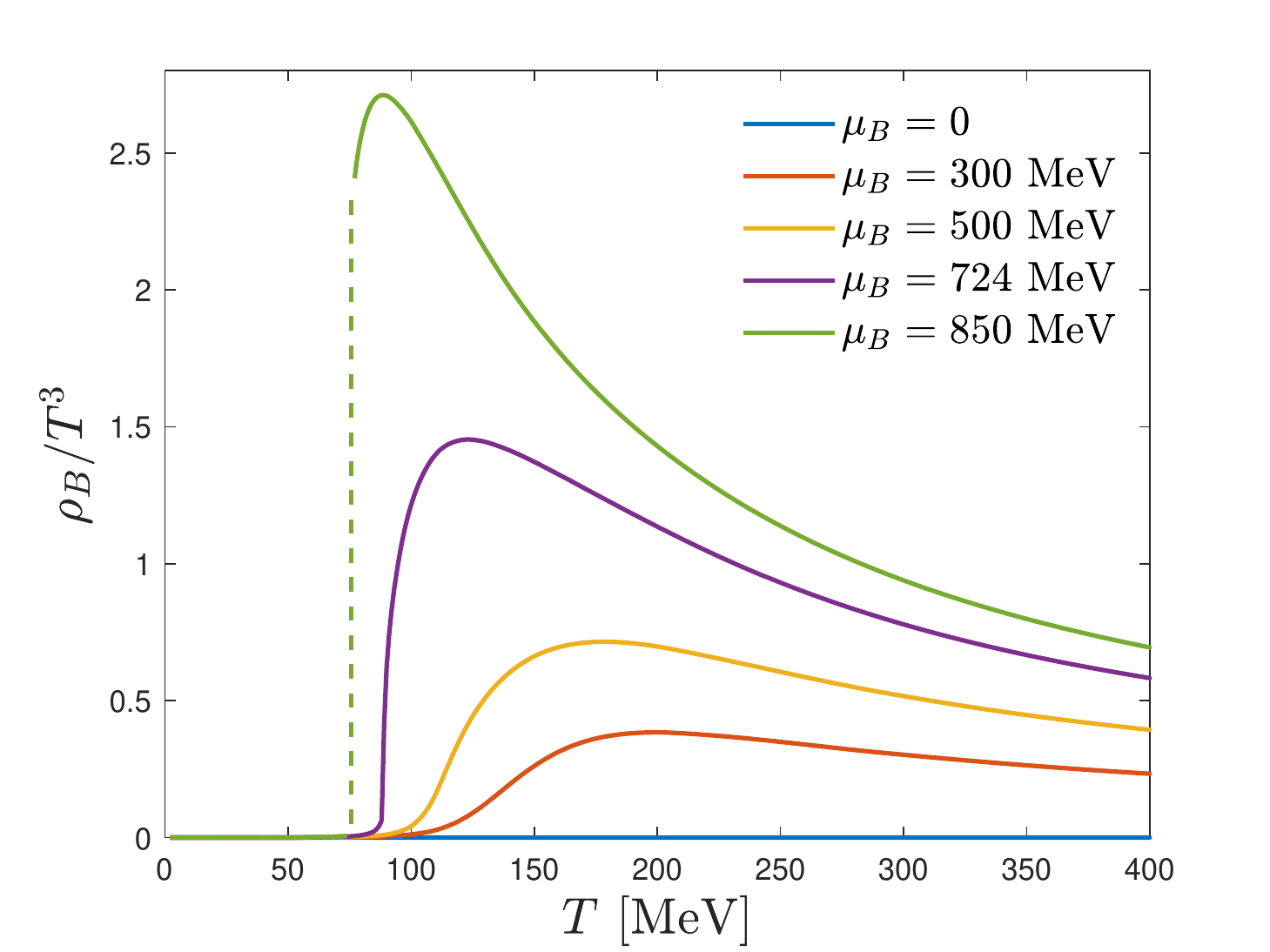}
    \newline
\centering
    \includegraphics[width=0.49\textwidth]{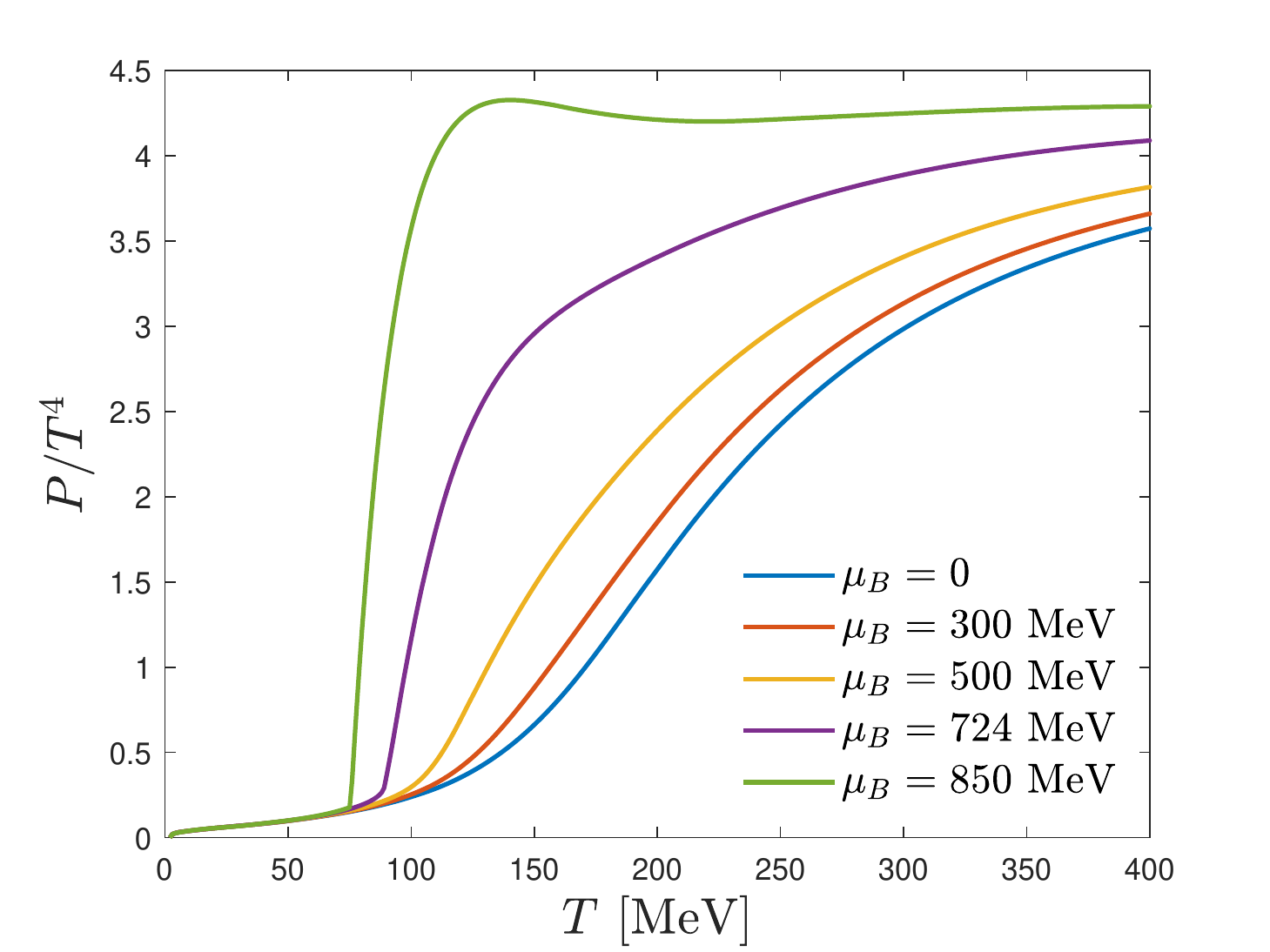}
    \includegraphics[width=0.49\textwidth]{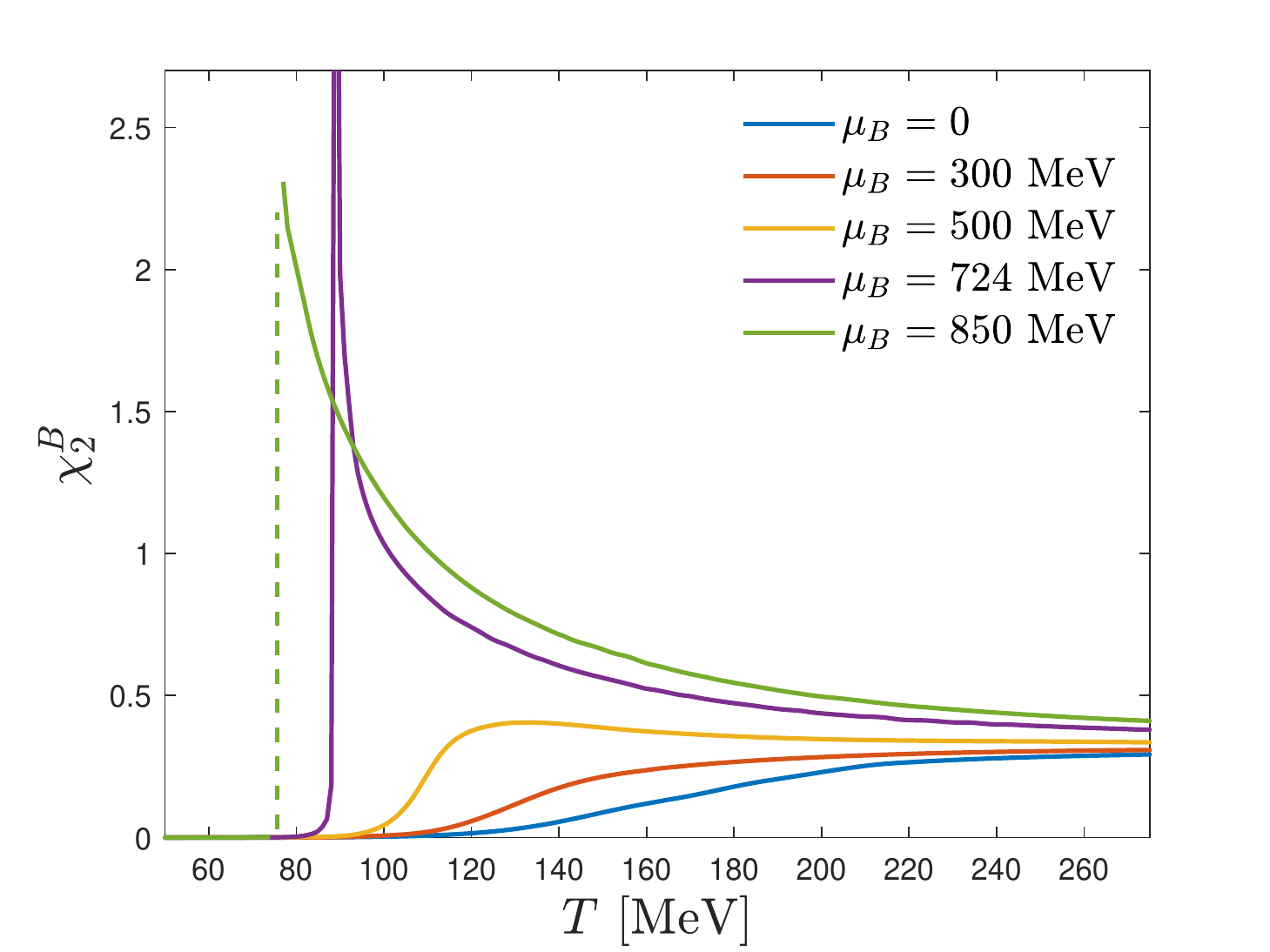}
    \caption{Equation of state as a function of the temperature for several values of the baryon chemical potential. The discontinuity exhibited by the entropy density, baryon number density and susceptibility at $\mu_{B}=850$ MeV (shown as a dashed line) corresponds to the line of first order phase transition. The divergence of $\chi_2^B$ at the critical point is clearly visible in the bottom right panel.}
    \label{fig:EoS_2D}
\end{figure*}

\begin{figure*}[htp!]
    \centering
    \includegraphics[width=0.49\textwidth]{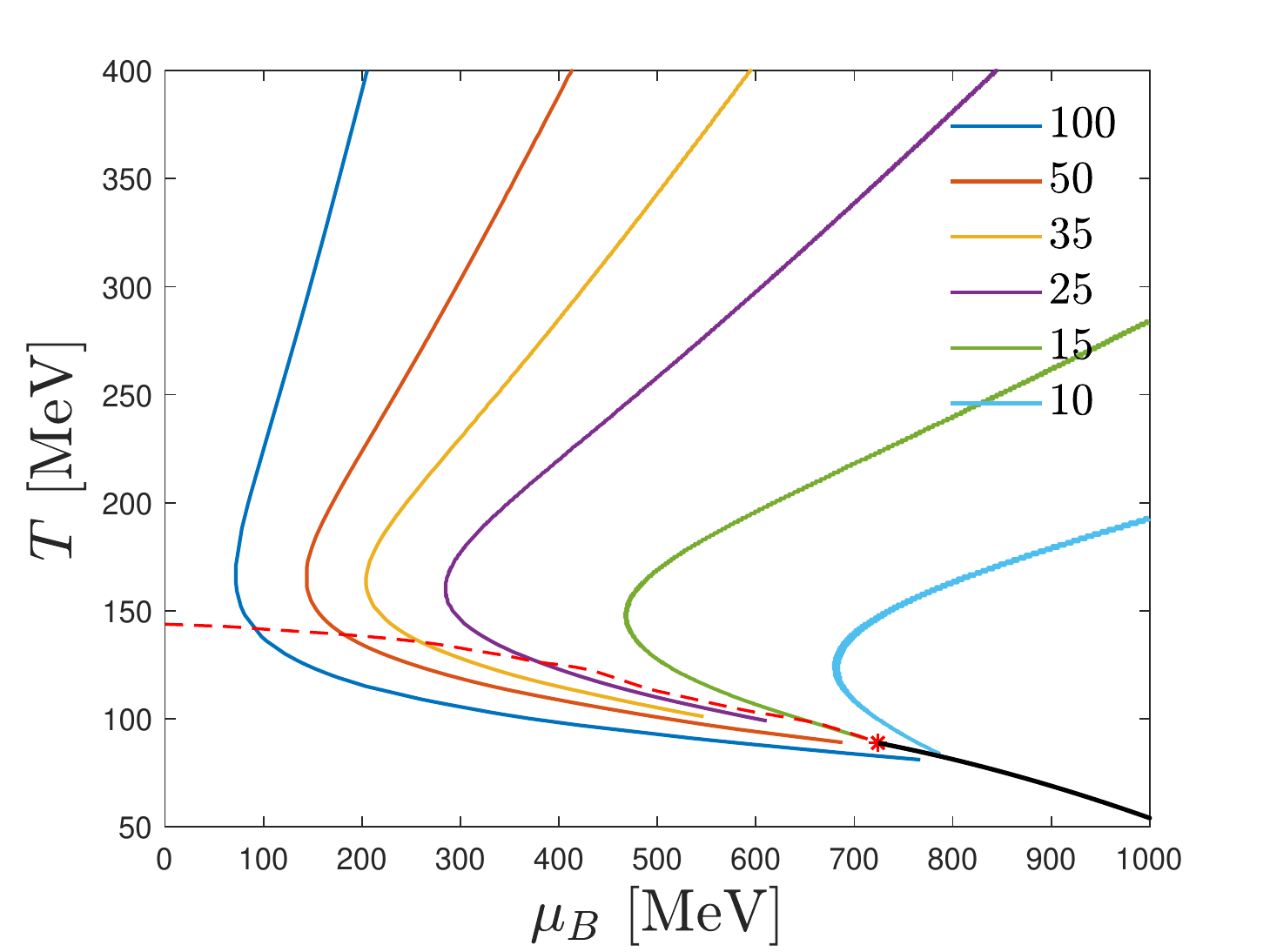}
    \includegraphics[width=0.49\textwidth]{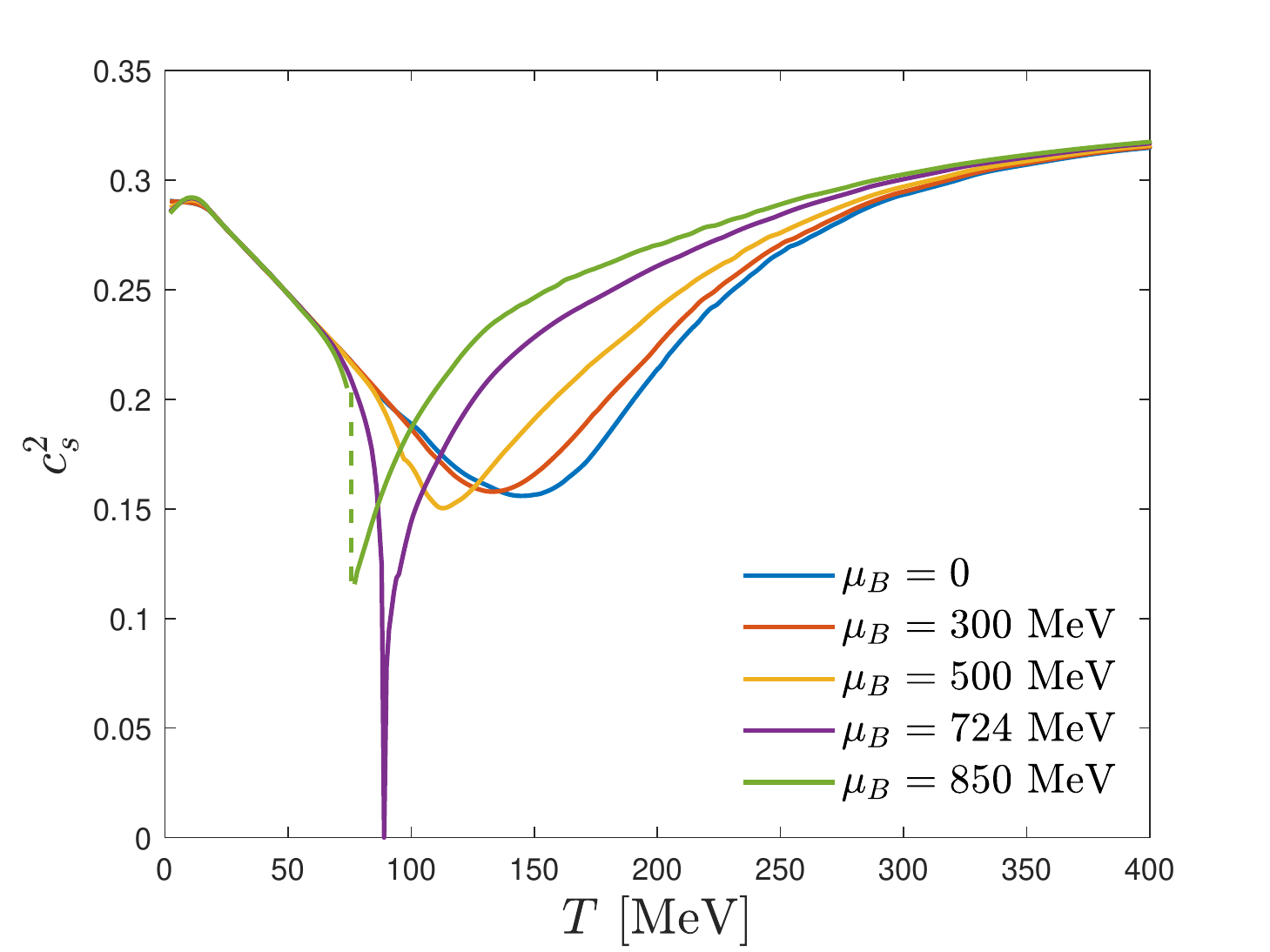}
    \caption{Left: Several isentropic trajectories in the $(T,\mu_B)$ plane. The red dashed line is the minimum of the square of the speed of sound that terminates at the critical point. The black curve is the first order transition line. Right: Square of the speed of sound at constant entropy per baryon number for different values of the chemical potential.}
    \label{fig:isentropes}
\end{figure*}

The critical point manifests itself in the first order derivatives of the pressure, namely the entropy and baryon density, where the pronounced gap shown by these state variables corresponds to the first-order phase transition line for $\mu_{B}>\mu_{B}^{\textrm{CEP}}$. In addition, $c_s^2$ exhibits a dip that becomes a zero at the CEP (which is a second-order phase transition point), and a discontinuity for $\mu_{B}>\mu_{B}^{\textrm{CEP}}$ along the first-order phase transition line, as expected from thermodynamic considerations. The location of the critical point is shown on the pressure surface as a red spot.


The dependence of the thermodynamic state variables on the temperature along lines of constant $\mu_B$ is presented in Fig.\ \ref{fig:EoS_2D}. There one can see the features mentioned above more explicitly such as the jump in entropy, baryon density, and $\chi_2^B$ after the first-order phase transition line is reached.  In \cite{Critelli:2017oub}, it was noted that the peak formation in $\chi_2^B$ may already indicate that a critical point is present at larger densities, and here this peak begins to happen around $\mu_B=416$ MeV. We see that our results appear to be consistent with this idea. One can also notice a peak formation in $P/T^4$ for large values of $\mu_B$. While at $\mu_B=0$ it is believed that $P/T^4$ should monotonically increase with $T$ \cite{Appelquist:1999vs}, non-monotonic behavior is expected to appear at nonzero $\mu_B$ and low $T$ \cite{Bedaque:2014sqa,Hippert:2021gfs}. Therefore, the non-monotonic behavior displayed by $P/T^4$ in the regime where $\mu_B/T >1$ may not be so surprising. It would be interesting to investigate this non-monotonic region further in a future work.

It is worth mentioning that, in order to obtain the $c_s^2$ at constant entropy per particle we calculate $c_{s}^{2}$ from Eq.\ (\ref{eq:c2s}) and it was necessary to remove the noise associated with the second order derivatives of the pressure. For a region up to $\mu_B\leq 600$ MeV, a SG filter was employed considering that the minimum of this observable is not very deep. However, due to the fact that the noise increases near the critical point for any observable in addition to the expected divergences in some derivatives, $\partial^{2}_{T}P$ and $\partial^{2}_{\mu_{B}}P$ for example, and the fact that at the critical point we expected a minimum, it was not possible to remove the noise without affecting the shape and features of the speed of sound.

Therefore, another approach was implemented to obtain the region of large chemical potential in front of the critical point. By obtaining the isentropic trajectories, one can also compute the speed of sound by taking the simple derivative $\partial P/\partial\epsilon$ along the isentropic paths. The lines of constant entropy over baryon density $s/\rho_{B}$ are relevant in the QCD phase diagram since they approximate the trajectories that the systems created in relativistic heavy ion collisions follow during their evolution when viscous effects are neglected. In fact, in the ideal case of vanishing viscosity, the quantity $s/\rho_{B}$ is conserved because the entropy generation is only caused by particle generation (although it has been shown that at large baryon densities and near a critical point, large deviations from this can be expected \cite{Dore:2020jye}).

The regions where the $c_s^2$ computed from Eq.\ (\ref{eq:c2s}) was heavily affected by the noise were removed and replaced with the information given by computing the same observable along the isentropic trajectories through an interpolation. Some isentropic lines are shown in the left panel of Fig.\ \ref{fig:isentropes}, along with the dependence of the $c_s^2$ with respect to the temperature for different values of the chemical potential (right panel).

\subsection{About the holographic model predictions at large $\mu_{B}$}

The results for the location of the CEP, and also for the thermodynamics in general, strongly depend on the choice for $V(\phi)$ and $f(\phi)$. Each different choice is in principle an effective holographic description of a different fluid at the boundary.

Ref.\ \cite{DeWolfe:2010he} laid down the foundations of the effective bottom-up holographic EMD approach which allows for a quantitative description of the strongly coupled QGP (and also of other kinds of strongly coupled physical systems, depending on how one fixes the free parameters of the model). However, at that time, two competing and incompatible lattice QCD results for the EOS had been proposed in the literature. The lattice data used in Ref.\ \cite{DeWolfe:2010he} to phenomenologically fix the profiles for $V(\phi)$ and $f(\phi)$ were later shown to be not the correct results for the QCD EOS (on a quantitative level). It is now well established in the lattice literature that the correct results for the QCD EOS and the 2nd order baryon susceptibility at zero baryon density are given in Refs.\ \cite{Borsanyi:2013bia,Bazavov:2014pvz,Bellwied:2015lba}. We have used these state-of-the-art lattice results to construct the EMD model of Ref.\ \cite{Critelli:2017oub}.

It remains to be determined whether other functional forms for $f(\phi)$ and $V(\phi)$ exist which provide a good description of the state-of-the-art lattice QCD results and might lead to a different location for the critical point and the line of first-order phase transition. This goes beyond the purpose of our present analysis, and will be investigated in future work.

We remark that the phenomenological reliability of a bottom-up holographic model, since the precise details of the dual field theory are unknown (contrary to top-down approaches), needs to be checked by directly comparing the predictions of the model under consideration against the target phenomenology which it is aimed to describe.

It is quite a common practice in the literature to fix the free parameters of an effective model to reproduce known features from experiments or first principle calculations, and then extrapolate the model predictions to regions which the fundamental theory cannot reach. A crucial requirement is that, once the free parameters are fixed, the model \textit{predictions} should effectively reproduce first principle results where they are available. In this regard, the effectiveness of the EMD model constructed in Ref.\ \cite{Critelli:2017oub}, and further analyzed in the present work, is currently unmatched in the literature (be it holographic or not).

We point out the following facts which strongly support this claim:

\begin{itemize}
    \item [i.] First, in Fig.\ 2 of Ref.\ \cite{Critelli:2017oub} we compared the predictions of the holographic EMD model for the finite baryon density EOS with the corresponding lattice QCD data from Ref.\ \cite{Bazavov:2017dus}, obtaining quantitative agreement all the way up to the highest values of $\mu_{B}$ reached in state-of-the-art lattice simulations at that time. Moreover, in Fig.\ 1 of Ref.\ \cite{Critelli:2017oub} we also compared the predictions of the EMD model for the $n$th-order baryon susceptibilities $\chi_{n}^{B}(T, \mu_{B} = 0)$ up to $n=6$, \footnote{Notice that $n=2$ at $\mu_B=0$ is \textit{not} a prediction of the EMD model, since the 2nd order baryon susceptibility at $\mu_B=0$ is used to fix the form of $f(\phi)$ in the EMD model.} with the corresponding lattice QCD data from Refs.\ \cite{Bazavov:2017dus, Bellwied:2015lba}, also obtaining quantitative agreement. In Fig.\ 1 of Ref.\ \cite{Critelli:2017oub} we further predicted the behavior for the eighth-order baryon susceptibility at vanishing chemical potential, $\chi_{8}^{B}(T,\mu_{B}=0)$, which at that time had not been evaluated on the lattice; some time later the lattice result for $\chi_{8}^{B}(T,\mu_{B}=0)$ was first calculated in Ref.\ \cite{Borsanyi:2018grb}, and the holographic prediction originally made in Ref.\ \cite{Critelli:2017oub} was compared in Fig.\ 1 of Ref.\ \cite{Rougemont:2018ivt} with the corresponding lattice result, again attaining quantitative agreement and further confirming the reliability of the present EMD model in the baryon dense regime of QCD.
    
    \item [ii.] Second, in the present manuscript we compare in Fig.\ \ref{fig:lattice_comparison} the predictions of the same holographic EMD model for the finite baryon density EOS with the most recent lattice QCD data obtained in Ref.\ \cite{Borsanyi:2021sxv}, which goes beyond the values of $\mu_{B}$ reached in the previous lattice simulations of Ref.\ \cite{Bazavov:2017dus} in 2017. These results are discussed in the manuscript, where again we see quantitative agreement between the EMD model and lattice results, although for the baryon charge density we see that the holographic prediction deviates from the latest lattice data in the high temperature regime for sufficiently high values of $\mu_{B}$, setting a first limitation for the phenomenological reliability of our model in the baryon dense regime at high temperatures (which is above the phase transition region in the phase diagram).
\end{itemize}

To the best of our knowledge, no other effective model currently available in the literature, besides the present holographic EMD model, has been able to successfully accomplish the thermodynamic tests we mentioned in items i) and ii) above.

\section{Conclusions and future directions}
\label{sec:conclusion}

In this work we presented our most recent results and predictions on the thermodynamics and phase diagram of strongly interacting QCD matter, obtained through a bottom-up non-conformal holographic approach. We significantly improved our numerical techniques, thus expanding our coverage in temperature and baryon density well beyond  our previous work \cite{Critelli:2017oub}. We were able to obtain the first-order phase transition line beyond the critical point for the first time out to $\mu_B\sim 1100$ MeV. A good agreement is found between our equation of state and the corresponding lattice QCD results available at intermediate densities. The equation of state obtained here, covering the first-order phase transition line and the critical end point in the $(T,\mu_B)$ plane, can be readily used in hydrodynamic simulations of the matter created in heavy-ion collisions. Furthermore, these results can be used to build a bridge to the high-density, low temperature region of the QCD phase diagram needed in the description of neutron star mergers \cite{Dexheimer:2020zzs}. We point out that in our model, we do not see $c_s^2$ ever surpassing the conformal limit of $c_s^2=1/3$, despite predictions from neutron star mergers that this may happen at $T=0$ and baryon densities above nuclear saturation density \cite{Bedaque:2014sqa,Alford:2015dpa,Ranea-Sandoval:2015ldr,Tews:2018kmu,Tews:2018iwm,McLerran:2018hbz,Jakobus:2020nxw,Annala:2019puf,Zhao:2020dvu,Tan:2020ics}. However, the holographic approach employed here is not expected to be a good guide for the behavior of strongly interacting matter at $T=0$. Rather, our model is expected to be useful at finite temperature and chemical potential where it is conceivable that QCD matter still displays near perfect fluidity. 

In order to directly connect these results to the STAR Beam Energy Scan at RHIC, our next steps will be to couple this to a hadron resonance gas EOS at low temperatures. This step is crucial because hydrodynamics must freeze-out into particles and those particles must match the exact hadron chemistry in the EOS at freeze-out, which is not possible in the holographic model since no explicit hadrons are used in its construction (moreover, hadron thermodynamics is not expected to be described by any classical gauge/gravity construction, since the pressure in the hadron gas phase is suppressed by a factor of $\sim N_c^2$ relatively to the pressure in the deconfined plasma phase in a large $N_c$ expansion, which would require quantum string loop corrections in the holographic duality to be properly accounted for). Then, this could be fed into a relativistic viscous hydrodynamics model with baryon conservation and it would be very interesting to perform calculations with the transport coefficients extracted from this same EMD model. Such a study has never been performed before and it would be the first of its kind to have both the EOS and transport coefficients in a hydrodynamic simulation taken from the same theoretical framework. Once that is completed, one could compare to spectra and flow harmonics at the beam energy scan.  Finally, we note that in our previous work \cite{Critelli:2017oub} we did make direct comparisons between baryon susceptibilities and STAR results for net-proton fluctuations, but performing such a study here is beyond the scope of the current work.

We also expect to report in the near future new results coming from the present holographic model, regarding the calculation of transport coefficients such as the bulk viscosity, the baryon and thermal conductivities, the baryon diffusion coefficient, the jet quenching parameter, the heavy quark drag force and the Langevin diffusion coefficients, all of them evaluated across the entire phase diagram covered in the present work, including the phase transition region.

It would be also interesting to see a model such as ours, which is consistent with lattice QCD results, to be applied in real time out-of-equilibrium studies such as those performed in  \cite{Casalderrey-Solana:2013aba,Casalderrey-Solana:2016xfq,Attems:2017zam,Attems:2018gou,Critelli:2018osu,Rougemont:2018ivt,Attems:2019yqn,Folkestad:2019lam,Attems:2020qkg}. Such study is, however, beyond the scope of the present work.

\section*{Acknowledgements}
This  material is based upon work supported by the National Science Foundation under grant no. PHY-1654219 and by the US-DOE Nuclear Science Grant No. DE-SC0020633,  US-DOE Office of Science, Office of Nuclear Physics, within the framework of the Beam Energy Scan Topical (BEST) Collaboration.  J.N. is partially supported by
the U.S. Department of Energy, Office of Science, Office for Nuclear Physics under Award
No. DE-SC0021301. R.R. acknowledges financial support by Universidade do Estado do Rio de Janeiro (UERJ) and Funda\c{c}\~{a}o Carlos Chagas de Amparo \`{a} Pesquisa do Estado do Rio de Janeiro (FAPERJ).

\bibliography{M335}

\end{document}